\documentclass[traditabstract]{aa}

\usepackage{natbib}
\bibpunct{(}{)}{;}{a}{}{,} 
\usepackage{amssymb}
\usepackage{url}
\usepackage{rotating,graphicx}
\usepackage{longtable}
\usepackage[usenames,dvipsnames]{color}
\newcommand{\urltt}[1]{\texttt{#1}}
\newcommand{\micron}{\mbox{$\mu$m}}
\newcommand{\msun}{\mbox{$M_\odot$}}
\newcommand{\lsun}{\mbox{$L_\odot$}}
\newcommand{\msunyr}{\mbox{\msun\ yr$^{-1}$}}

\newcommand{\kms}{\mbox{km~s$^{-1}$}}

\newcommand{\grasil}{\sc Grasil}
\newcommand{\magphys}{\sc Magphys}
\newcommand{\cmg}{\mbox{cm}^2\,\mbox{g}^{-1}}
\newcommand{\grb}{GRB\,111005A}

\newlength{\propwidth}
\begin{document}

 \title{The second closest gamma-ray burst: sub-luminous  GRB\,111005A  with no supernova in a super-solar metallicity environment 
 }
 
\titlerunning{The second closest GRB: 111005A}
\authorrunning{Micha{\l}owski et al.}

\author{Micha{\l}~J.~Micha{\l}owski\inst{\ref{inst:poz},\ref{inst:roe}}
\and
Dong~Xu\inst{\ref{inst:dark},\ref{inst:bei}} 
\and
Jamie~Stevens\inst{\ref{inst:atca}} 
\and
Andrew Levan\inst{\ref{inst:war}} 
\and
Jun~Yang\inst{\ref{inst:ons},\ref{inst:jive}}
\and
Zsolt~Paragi\inst{\ref{inst:jive}} 
\and
Atish~Kamble\inst{\ref{inst:cfa}} 
\and
An-Li Tsai\inst{\ref{inst:poz}}
\and
Helmut~Dannerbauer\inst{\ref{inst:iac},\ref{inst:lag},\ref{inst:wien}} 
\and 
Alexander J.~van der Horst\inst{\ref{inst:wash}} 
\and
Lang Shao\inst{\ref{inst:lang},\ref{inst:nan}} 
\and
David Crosby\inst{\ref{inst:roe}} 
\and
Gianfranco~Gentile\inst{\ref{inst:gent},\ref{inst:brus}}
\and
Elizabeth Stanway\inst{\ref{inst:war}} 
\and
Klaas Wiersema\inst{\ref{inst:lest},\ref{inst:war}}
\and 
Johan P.~U.~Fynbo\inst{\ref{inst:dark}} 
\and
Nial~R.~Tanvir\inst{\ref{inst:lest}}
\and
Peter~Kamphuis\inst{\ref{inst:bochum}}
\and
Michael~Garrett\inst{\ref{inst:jodrell},\ref{inst:leiden}} 
\and
Przemys\l aw Bartczak\inst{\ref{inst:poz}}
	}

\institute{
Astronomical Observatory Institute, Faculty of Physics, Adam Mickiewicz University, ul.~S{\l}oneczna 36, 60-286 Pozna{\'n}, Poland, 
\label{inst:poz}
{\tt mj.michalowski@gmail.com}
\and
Scottish Universities Physics Alliance (SUPA), Institute for Astronomy, University of Edinburgh, Royal Observatory, Blackford Hill, Edinburgh, EH9 3HJ, UK, 
\label{inst:roe}
\and
Dark Cosmology Centre, Niels Bohr Institute, University of Copenhagen, Juliane Maries Vej 30, DK-2100 Copenhagen \O, Denmark  \label{inst:dark}
\and
National Astronomical Observatories, Chinese Academy of Sciences, Beijing 100012, China \label{inst:bei}
\and
CSIRO Astronomy and Space Science, Locked Bag 194, Narrabri NSW 2390, Australia \label{inst:atca}
\and
Department of Physics, University of Warwick, Coventry, CV4 7AL, UK \label{inst:war}
\and
Department of Earth and Space Sciences, Chalmers University of Technology, Onsala Space Observatory, SE-43992 Onsala, Sweden \label{inst:ons}
\and
Joint Institute for VLBI ERIC, Postbus 2, NL-7990 AA Dwingeloo, the Netherlands \label{inst:jive}
\and
Harvard-Smithsonian Center for Astrophysics, 60 Garden St., Cambridge, MA 02138 \label{inst:cfa}
\and
Instituto de Astrofisica de Canarias (IAC), E-38205 La Laguna, Tenerife, Spain \label{inst:iac}
\and
Universidad de La Laguna, Dpto. Astrofisica, E-38206 La Laguna, Tenerife, Spain \label{inst:lag}
\and
Universit\"{a}t Wien, Institut f\"{u}r Astrophysik, T\"{u}rkenstra{\ss}e 17, 1180 Wien, Austria \label{inst:wien}
\and
Department of Physics, The George Washington University, 725 21st Street NW, Washington, DC 20052, USA \label{inst:wash}
\and
Department of Space Sciences and Astronomy, Hebei Normal University, Shijiazhuang 050024, China \label{inst:lang}
\and
Key Laboratory of Dark Matter and Space Astronomy, Purple Mountain Observatory, Chinese Academy of Sciences, Nanjing 210008, China \label{inst:nan}
\and
Sterrenkundig Observatorium, Universiteit Gent, Krijgslaan 281-S9, 9000, Gent, Belgium  \label{inst:gent}
\and
Department of Physics and Astrophysics, Vrije Universiteit Brussel, Pleinlaan 2, 1050 Brussels, Belgium \label{inst:brus}
\and
Department of Physics and Astronomy, University of Leicester, Leicester LE1 7RH \label{inst:lest}
\and
Astronomisches Institut, Ruhr-Universit\"at Bochum, Universit\"atsstrasse 150, 44801 Bochum, Germany \label{inst:bochum}
\and
Jodrell Bank Centre for Astrophysics, School of Physics \& Astronomy, The University of Manchester, Alan 
Turing Building, Oxford Road, Manchester, M13 9PL, UK \label{inst:jodrell}
\and
Leiden Observatory, University of Leiden, PO Box 9513, NL-2300 RA Leiden, the Netherlands \label{inst:leiden}
}

\abstract{
We report the detection of the radio afterglow of a long gamma-ray burst (GRB) 111005A at $5$--$345\,$GHz, including the very long baseline interferometry observations with the positional error of $0.2$ mas. The afterglow position is coincident with the disk of a galaxy ESO 580-49 at $z= 0.01326$ ($\sim1\arcsec$ from its center), which makes {\grb} the second closest GRB known to date, after GRB\,980425. 
The radio afterglow of GRB\,111005A was  an order of magnitude less luminous than those of local low-luminosity GRBs, and obviously than those of cosmological GRBs.   The radio flux was approximately constant and then experienced an unusually rapid decay a month after the GRB explosion. 
Similarly to only two other GRBs, we did not find the associated supernovae (SN), despite deep near- and mid-infrared observations 1--9 days after the GRB explosion, reaching $\sim20$ times fainter than other SNe associated with GRBs. Moreover, we measured twice solar metallicity for the GRB location. The low $\gamma$-ray and radio luminosities, rapid decay, lack of a SN, and super-solar metallicity suggest that  {\grb} represents a different rare class of GRBs than typical core-collapse events.  We modelled the spectral energy distribution of the {\grb} host finding that it is a dwarf, moderately star-forming galaxy, similar to the host of GRB\,980425. The existence of two local GRBs in  such galaxies is still consistent with the hypothesis that the GRB rate is proportional to the cosmic star formation rate (SFR) density, but suggests that the GRB rate is biased towards low SFRs. Using the far-infrared detection of ESO 580-49, we conclude that the hosts of both GRBs 111005A and 980425 exhibit lower dust content than what would be expected from their stellar masses and optical colours.
}


\maketitle

\section{Introduction}
\label{sec:intro}

\begin{figure*}
\begin{center}
\includegraphics[width=\textwidth,viewport=85 745 530 830,clip]{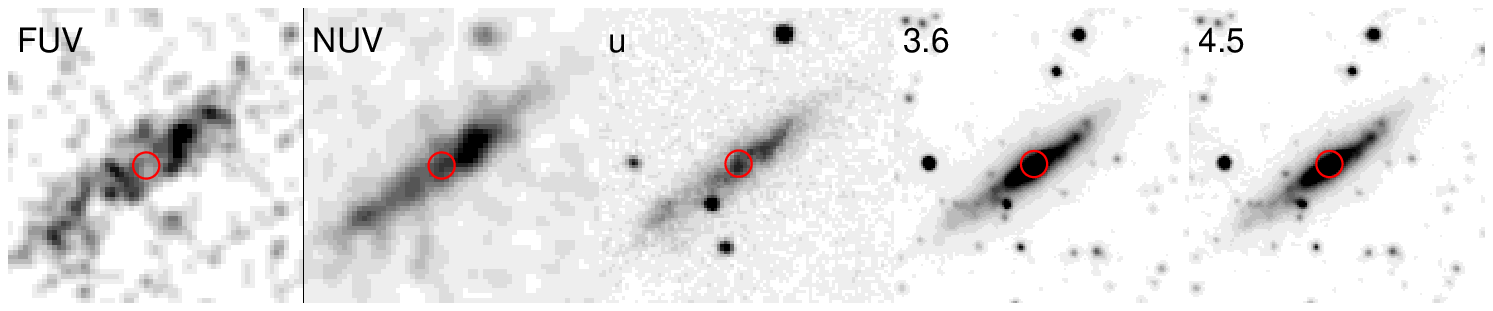}
\end{center}
\caption{Mosaic of the images of  the {\grb} host. The images are from {\it GALEX} (FUV and NUV),  {\it Swift}/UVOT ($u$-band) and {\it Spitzer} \citep{levan11spitzer}. North is up and east is to the left. Each panel is  $90''\times90''$  ($24\mbox{ kpc}\times24\mbox{ kpc}$). 
 The {\it red circle}  shows the VLBA position. 
}
\label{fig:im}
\end{figure*}

\begin{figure}
\begin{center}
\includegraphics[angle=0,width=0.35\textwidth,clip,viewport=180 10 410 840]{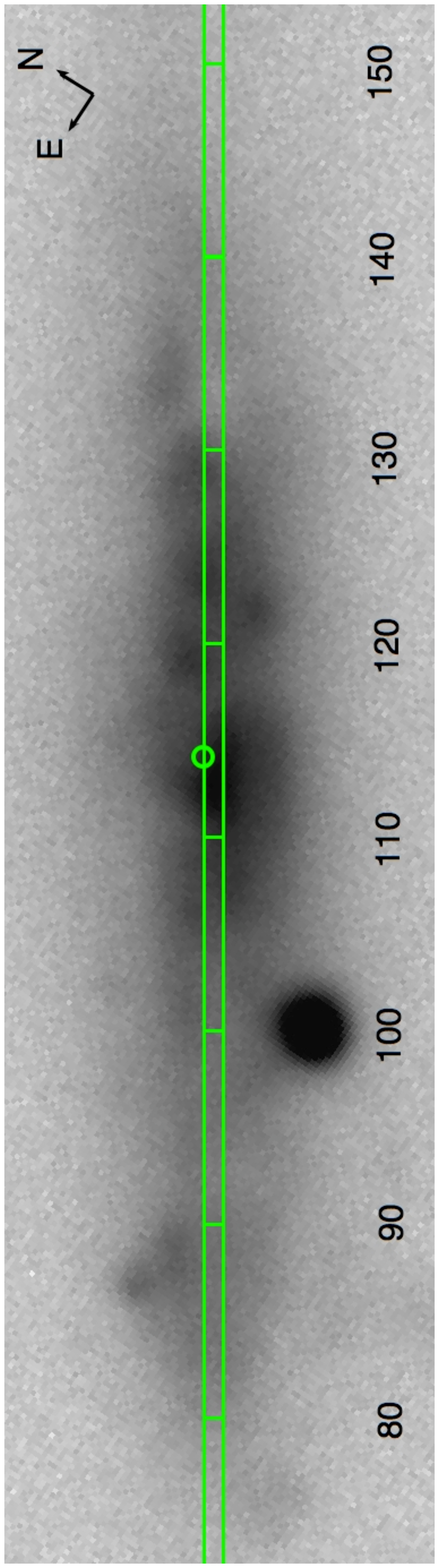}
\end{center}
\caption{Position of the WHT slit with regions marked by their distance from its beginning. The BPT diagnostic and metallicities of these regions are shown in Fig.~\ref{fig:bpt} and \ref{fig:metal}, respectively. The circle denotes the afterglow VLBA position.}
\label{fig:slit}
\end{figure}

\begin{figure*}
\begin{center}
\includegraphics[width=\textwidth,viewport=85 745 530 830,clip]{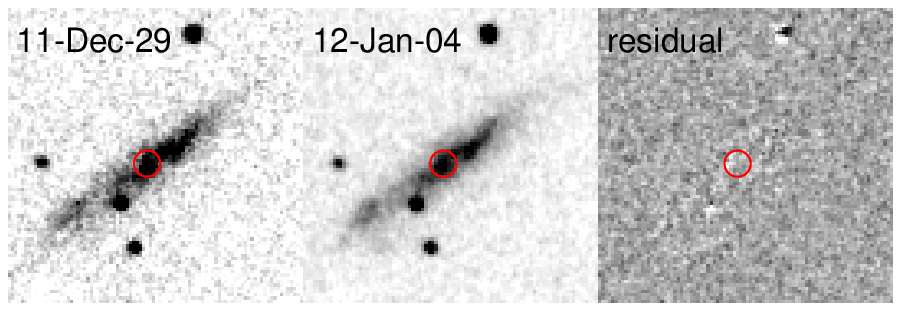}
\end{center}
\caption{{\it Swift}/UVOT $u$-band images of  the {\grb} host taken at two epochs and the result of the subtraction showing no variable source. North is up and east is to the left. Each panel is  $90\arcsec\times90\arcsec$  ($24\mbox{ kpc}\times24\mbox{ kpc}$). 
 The {\it red circle}  shows the VLBA position. 
}
\label{fig:imu}
\end{figure*}

Long (duration $>2$ s) gamma ray-burst (GRBs) have been shown to be collapses of very massive stars (e.g.~\citealt{hjorthnature,stanek}; see a review in \citealt{hjorthsn}), and because of very short main-sequence lifetimes of such stars, GRBs are expected to trace galaxies with on-going star-formation \citep[but see][]{rossi14}. This could potentially be used as a tool to study cosmic star formation rate (SFR) density, but requires prior understanding of GRBs and their host galaxies.
Most of GRBs reside at $z\sim2$--$3$ \citep{jakobsson06,jakobsson12,fynbo09,greiner11,hjorth12,kruhler12b,salvaterra12,perley16} and there are only a few examples of low-$z$ GRBs. Hence, the GRB rate and properties at low-$z$ is very poorly constrained.

Low-$z$ GRBs and their hosts provide an opportunity to study their properties at the level of details inaccessible for more distant examples. For example, the local environments of GRBs, characterised with high-resolution observations, provide constraints on the age, mass and the explosion mechanism of the GRB progenitor  \citep{castroceron06,ostlin08,leloudas11,levan14,arabsalmani15b,greiner16}.
Moreover, low-$z$ GRBs are promising candidates of the detection of non-electromagnetic signals, like gravitational waves and neutrinos.

Similarly, the radio/submm observations of  afterglows of low-$z$ GRBs (see \citealt{weiler02} for a review and \citealt{deugartepostigo12} for a compilation) allow the measurements of the physical conditions of the explosion and the surrounding circumburst medium \citep{soderberg04,frail05,taylor05,vanderhorst05,vanderhorst08,vanderhorst14,margutti13}, and even of the size of the expanding ejecta, if very long baseline interferometry (VLBI) observations are available  \citep[][]{taylor04,pihlstrom07}.

When it comes to the host galaxies, only thirteen of them have been detected in the far-infrared (but see \citealt{perley17b}): those of GRB 
980425 \citep{lefloch12,michalowski14}, 
980613, 020819B, 051022, 070306, 080207, 080325, 090417B \citep{hunt14,hatsukade14,schady14},
010222 \citep{frail},
000210, 000418 \citep{berger,tanvir},
031203 \citep{watson11, symeonidis14},
and 080607 \citep{wang12}.
Hence,  we still do not posses a significant sample of GRB hosts whose dust emission can be studied, and this is where low-$z$ GRBs can be useful.

{\grb} triggered the Burst Alert Telescope (BAT; \citealt{bat}) on board of the {\it Swift} satellite \citep{swift} at 08:05:14 UT on 2011 Oct 5. The burst was localised at 14:53:08, $-$19:43:48 with 90\% error circle of $3'$ \citep{saxton11gcn}, further revised to $2.1'$ \citep{barthelmy11gcn}. The duration of $26\pm7$ sec \citep{barthelmy11gcn} classifies it in the long GRB category \citep{kouveliotou93}.  \citet{barthelmy11gcn} reported the burst's power law spectral index of $2.03\pm0.27$, the fluence in the $15$--$150$ keV band of $(6.2\pm1.1)\times10^{-7}\,\mbox{erg cm}^2$ and the peak photon flux in this band of  $1.1\pm0.3\,\mbox{ph cm}^{-2}\, \mbox{s}^{-1}$. At the time of the burst the Sun was close to its position, so no X-ray or optical observations in the early stages were possible. Near-infrared images taken during twilight and close to the horizon did not reveal any variable source \citep{levan11gcn2,nardini11gcn,malesani11gcn,motohara11gcn}.  The potential association of {\grb} to a galaxy ESO 580-49 at $z= 0.01326$ was suggested by \citet{levan11gcn}, whereas \citet{zauderer11gcn} detected a radio source coincident with this galaxy (EVLA-S1 at 14:53:07.78, $-$19:44:12.2). The association of the GRB and this local galaxy is confirmed by our multi-facility campaign presented in this paper and reported initially in  \citet{xu11gcn,xu11gcn2} and \citet{michalowski11gcn}.

The objectives of these papers are: {\it i)} report the discovery and the confirmation of the low redshift of {\grb},  {\it ii)} determine the nature of this GRB, and {\it iii)} study its host galaxy in the context of other GRB hosts and of local star-forming galaxies. 

We use a cosmological model with $H_0=70$ km s$^{-1}$ Mpc$^{-1}$,  $\Omega_\Lambda=0.7$, and $\Omega_m=0.3$, so {\grb} at $z= 0.01326$ is at a luminosity distance of   57.4 Mpc and $1\arcsec$ corresponds to 271 pc at its redshift. We also assume the 
\citet{chabrier03} 
initial mass function (IMF), to which all SFR and stellar masses were converted (by dividing by 1.8) if given originally assuming the \citet{salpeter} IMF.

\section{Data}
\label{sec:data}

\begin{table*}
\caption{The results of the afterglow observations of GRB 111005A \label{tab:after}   }
\centering
\begin{tabular}{lcccccl}
\hline\hline
Date\tablefootmark{a} & $\Delta t$\tablefootmark{b} & Freq.  & Flux  & Beam\tablefootmark{c} & PA\tablefootmark{d} & Instrument\tablefootmark{e}\\
(yyyy-mm-dd-hh.h)                & (days)           & (GHz)  & (mJy) &   ($\arcsec$) & (deg)\\
\hline
2011-10-05-23.49514\tablefootmark{f} & \phantom{1}\phantom{1}0.64200 $\pm$ \phantom{1}0.00496 & \phantom{1}\phantom{1}5.8 & \phantom{1}\phantom{$-$}0.2902 $\pm$ \phantom{1}0.0109 & $26.73 \times 12.39$ & 30 & VLA/D \\
2011-10-05-23.99514\tablefootmark{f} & \phantom{1}\phantom{1}0.66283 $\pm$ \phantom{1}0.00496 & \phantom{1}\phantom{1}5.8 & \phantom{1}\phantom{$-$}0.2684 $\pm$ \phantom{1}0.0137 & $29.16 \times 11.23$ & 36 & VLA/D \\
2011-10-06-22.03056\tablefootmark{f} & \phantom{1}\phantom{1}1.58097 $\pm$ \phantom{1}0.00495 & \phantom{1}\phantom{1}5.8 & \phantom{1}\phantom{$-$}0.2999 $\pm$ \phantom{1}0.0126 & $22.87 \times 12.77$ & 2 & VLA/D \\
2011-10-07-05.96056 & \phantom{1}\phantom{1}1.91139 $\pm$ \phantom{1}0.02704 & \phantom{1}\phantom{1}9.0 & \phantom{1}$-$0.0527 $\pm$ \phantom{1}0.0640 & $4.89 \times 0.58$ & 16 & ATCA/H75 \\
2011-10-07-05.96056 & \phantom{1}\phantom{1}1.91139 $\pm$ \phantom{1}0.02704 & \phantom{1}\phantom{1}5.5 & \phantom{1}$-$0.0267 $\pm$ \phantom{1}0.0380 & $8.07 \times 0.95$ & 16 & ATCA/H75 \\
2011-10-08-07.88569 & \phantom{1}\phantom{1}2.99160 $\pm$ \phantom{1}0.03350 & \phantom{1}18.0 & \phantom{1}\phantom{$-$}1.9100 $\pm$ \phantom{1}0.0700 & $72.67 \times 18.73$ & -53 & ATCA/H75 \\
2011-10-08-21.39208\tablefootmark{f} & \phantom{1}\phantom{1}3.55437 $\pm$ \phantom{1}0.00496 & \phantom{1}\phantom{1}5.8 & \phantom{1}\phantom{$-$}0.2857 $\pm$ \phantom{1}0.0115 & $19.65 \times 10.68$ & 6 & VLA/D \\
2011-10-10-03.78750 & \phantom{1}\phantom{1}4.82084 $\pm$ \phantom{1}0.04111 & \phantom{1}34.0 & \phantom{1}\phantom{$-$}2.3100 $\pm$ \phantom{1}0.0400 & $15.33 \times 9.60$ & 82 & ATCA/H75 \\
2011-10-10-05.81542 & \phantom{1}\phantom{1}4.90534 $\pm$ \phantom{1}0.02920 & \phantom{1}18.0 & \phantom{1}\phantom{$-$}1.4100 $\pm$ \phantom{1}0.0600 & $33.51 \times 18.51$ & -74 & ATCA/H75 \\
2011-10-10-07.24181 & \phantom{1}\phantom{1}4.96477 $\pm$ \phantom{1}0.01832 & \phantom{1}94.0 & \phantom{$-$}14.8000 $\pm$ \phantom{1}0.4000 & $11.20 \times 3.78$ & -55 & ATCA/H75 \\
2011-10-10-20.17956 & \phantom{1}\phantom{1}5.50385 $\pm$ \phantom{1}0.01870 & 345.0 & \phantom{$-$}14.0000 $\pm$ 13.0000 & $17.92 $ & $\cdots$  & APEX \\
2011-10-12-03.83972 & \phantom{1}\phantom{1}6.82302 $\pm$ \phantom{1}0.05812 & \phantom{1}18.0 & \phantom{1}\phantom{$-$}1.2100 $\pm$ \phantom{1}0.0400 & $29.07 \times 17.69$ & -83 & ATCA/H75 \\
2011-10-13-04.65347 & \phantom{1}\phantom{1}7.85693 $\pm$ \phantom{1}0.04134 & \phantom{1}18.0 & \phantom{1}\phantom{$-$}1.4700 $\pm$ \phantom{1}0.0500 & $29.70 \times 18.37$ & -81 & ATCA/H75 \\
2011-10-14-03.55847 & \phantom{1}\phantom{1}8.81130 $\pm$ \phantom{1}0.02307 & \phantom{1}34.0 & \phantom{1}\phantom{$-$}1.9400 $\pm$ \phantom{1}0.0700 & $15.04 \times 9.71$ & -77 & ATCA/H75 \\
2011-10-14-06.93319 & \phantom{1}\phantom{1}8.95192 $\pm$ \phantom{1}0.04729 & \phantom{1}18.0 & \phantom{1}\phantom{$-$}1.8600 $\pm$ \phantom{1}0.0600 & $42.14 \times 18.95$ & -62 & ATCA/H75 \\
2011-10-16-02.81389 & \phantom{1}10.78028 $\pm$ \phantom{1}0.03279 & \phantom{1}18.0 & \phantom{1}\phantom{$-$}1.8500 $\pm$ \phantom{1}0.0300 & $66.57 \times 19.01$ & -54 & ATCA/H75 \\
2011-10-17-04.34625 & \phantom{1}11.84413 $\pm$ \phantom{1}0.05470 & \phantom{1}18.0 & \phantom{1}\phantom{$-$}1.5100 $\pm$ \phantom{1}0.0400 & $41.34 \times 19.70$ & -61 & ATCA/H75 \\
2011-10-18-01.25875\tablefootmark{g} & \phantom{1}12.71548 $\pm$ \phantom{1}0.57813 & \phantom{1}\phantom{1}5.0 & \phantom{1}\phantom{$-$}0.0000 $\pm$ \phantom{1}0.1100 & $0.02000 \times 0.00700$ & 0 & EVN \\
2011-10-18-01.25875\tablefootmark{g} & \phantom{1}12.71548 $\pm$ \phantom{1}0.57813 & \phantom{1}\phantom{1}5.0 & \phantom{1}\phantom{$-$}0.4400 $\pm$ \phantom{1}0.1000 & $34.51 \times 3.86$ & 2 & WSRT \\
2011-10-20-04.74153 & \phantom{1}14.86060 $\pm$ \phantom{1}0.04148 & \phantom{1}34.0 & \phantom{1}\phantom{$-$}1.7600 $\pm$ \phantom{1}0.0700 & $17.72 \times 9.68$ & -76 & ATCA/H75 \\
2011-10-20-06.70014 & \phantom{1}14.94220 $\pm$ \phantom{1}0.03997 & \phantom{1}18.0 & \phantom{1}\phantom{$-$}1.6600 $\pm$ \phantom{1}0.0300 & $46.31 \times 18.96$ & -59 & ATCA/H75 \\
2011-10-21-20.03319 & \phantom{1}16.49775 $\pm$ \phantom{1}0.12638 & \phantom{1}15.0 & \phantom{1}\phantom{$-$}0.6660 $\pm$ \phantom{1}0.1500 & $0.00131 \times 0.00049$ & -6 & VLBA \\
2011-10-24-03.27972 & \phantom{1}18.79969 $\pm$ \phantom{1}0.03161 & \phantom{1}34.0 & \phantom{1}\phantom{$-$}1.7600 $\pm$ \phantom{1}0.0400 & $14.59 \times 10.14$ & -76 & ATCA/H75 \\
2011-10-24-04.78639 & \phantom{1}18.86247 $\pm$ \phantom{1}0.03081 & \phantom{1}18.0 & \phantom{1}\phantom{$-$}1.4600 $\pm$ \phantom{1}0.0500 & $34.18 \times 17.98$ & -75 & ATCA/H75 \\
2011-11-02-05.27958 & \phantom{1}27.88302 $\pm$ \phantom{1}0.02209 & \phantom{1}34.0 & \phantom{1}\phantom{$-$}2.4000 $\pm$ \phantom{1}0.2000 & $5.60 \times 0.17$ & 20 & ATCA/750C \\
2011-11-02-06.52667 & \phantom{1}27.93498 $\pm$ \phantom{1}0.02525 & \phantom{1}18.0 & \phantom{1}\phantom{$-$}1.5900 $\pm$ \phantom{1}0.0800 & $6.17 \times 0.42$ & 36 & ATCA/750C \\
2011-11-08-05.10194 & \phantom{1}33.87561 $\pm$ \phantom{1}0.02859 & \phantom{1}18.0 & \phantom{1}\phantom{$-$}1.4300 $\pm$ \phantom{1}0.0500 & $8.87 \times 0.40$ & 23 & ATCA/EW367 \\
2011-11-24-10.50000 & \phantom{1}50.10053 $\pm$ \phantom{1}0.08681 & \phantom{1}\phantom{1}5.0 & \phantom{1}\phantom{$-$}0.0000 $\pm$ \phantom{1}0.0350 & $0.02570 \times 0.00660$ & -25 & EVN \\
2011-12-05-19.68472 & \phantom{1}61.48323 $\pm$ \phantom{1}0.01591 & \phantom{1}18.0 & \phantom{1}\phantom{$-$}0.0967 $\pm$ \phantom{1}0.0530 & $21.50 \times 0.76$ & -37 & ATCA/6A \\
2011-12-14-19.68361 & \phantom{1}70.48318 $\pm$ \phantom{1}0.01565 & \phantom{1}34.0 & \phantom{1}\phantom{$-$}0.1187 $\pm$ \phantom{1}0.0660 & $12.14 \times 0.40$ & -48 & ATCA/6A \\
2011-12-14-20.63167 & \phantom{1}70.52269 $\pm$ \phantom{1}0.01457 & \phantom{1}18.0 & \phantom{1}\phantom{$-$}0.0321 $\pm$ \phantom{1}0.0370 & $24.56 \times 0.76$ & -46 & ATCA/6A \\
2011-12-10-08.14208\tablefootmark{g} & \phantom{1}66.00229 $\pm$ \phantom{1}4.53497 & \phantom{1}18.0 & \phantom{1}\phantom{$-$}0.0000 $\pm$ \phantom{1}0.0250 & $23.03 \times 0.76$ & -42 & ATCA/6A \\
2011-12-23-07.68361\tablefootmark{g} & \phantom{1}78.98318 $\pm$ \phantom{1}8.51565 & \phantom{1}34.0 & \phantom{1}\phantom{$-$}0.0000 $\pm$ \phantom{1}0.0390 & $12.14 \times 0.40$ & -46 & ATCA/6A \\
2011-12-18-18.48250 & \phantom{1}74.43314 $\pm$ \phantom{1}0.02884 & \phantom{1}18.0 & \phantom{1}$-$0.1860 $\pm$ \phantom{1}0.1150 & $12.33 \times 0.76$ & -50 & ATCA/6A \\
2011-12-19-10.46528 & \phantom{1}75.09909 $\pm$ \phantom{1}0.08299 & \phantom{1}\phantom{1}5.0 & \phantom{1}\phantom{$-$}0.1166 $\pm$ \phantom{1}0.0597 & $9.75 \times 3.22$ & 2 & WSRT \\
2011-12-23-10.45139 & \phantom{1}79.09851 $\pm$ \phantom{1}0.06215 & \phantom{1}\phantom{1}8.3 & \phantom{1}\phantom{$-$}0.0000 $\pm$ \phantom{1}0.3500 & $37.48 \times 2.00$ & 4 & WSRT \\
2011-12-26-06.40000 & \phantom{1}81.92970 $\pm$ \phantom{1}0.02917 & \phantom{1}94.5 & \phantom{1}\phantom{$-$}0.0000 $\pm$ \phantom{1}0.0800 & $12.20 \times 4.60$ & 157 & PdBI/D \\
2011-12-28-07.90694 & \phantom{1}83.99249 $\pm$ \phantom{1}0.07292 & \phantom{1}\phantom{1}2.3 & \phantom{1}\phantom{$-$}0.0000 $\pm$ \phantom{1}0.1210 & $65.84 \times 5.85$ & 3 & WSRT \\
2012-01-03-09.77361 & \phantom{1}90.07027 $\pm$ \phantom{1}0.07083 & \phantom{1}\phantom{1}5.0 & \phantom{1}\phantom{$-$}0.1869 $\pm$ \phantom{1}0.0697 & $68.16 \times 3.29$ & 2 & WSRT \\
2012-01-10-21.49694 & \phantom{1}97.55874 $\pm$ \phantom{1}0.02175 & \phantom{1}18.0 & \phantom{1}\phantom{$-$}0.1500 $\pm$ \phantom{1}0.0290 & $17.22 \times 0.76$ & -40 & ATCA/6A \\
2012-01-11-20.11750 & \phantom{1}98.50126 $\pm$ \phantom{1}0.01551 & \phantom{1}34.0 & \phantom{1}$-$0.0214 $\pm$ \phantom{1}0.0430 & $12.14 \times 0.40$ & -44 & ATCA/6A \\
2013-07-21-23.25833\tablefootmark{f}\tablefootmark{g} & 655.63213 $\pm$ \phantom{1}3.63299 & \phantom{1}\phantom{1}1.4 & \phantom{1}\phantom{$-$}0.2200 $\pm$ \phantom{1}0.0200 & $28.72 \times 4.35$ & 4 & ATCA/6A \\
2013-07-21-23.25833\tablefootmark{f}\tablefootmark{g} & 655.63213 $\pm$ \phantom{1}3.63299 & \phantom{1}\phantom{1}1.9 & \phantom{1}\phantom{$-$}0.1800 $\pm$ \phantom{1}0.0200 & $17.70 \times 3.28$ & 6 & ATCA/6A \\
2013-07-21-23.25833\tablefootmark{f}\tablefootmark{g} & 655.63213 $\pm$ \phantom{1}3.63299 & \phantom{1}\phantom{1}2.4 & \phantom{1}\phantom{$-$}0.1500 $\pm$ \phantom{1}0.0200 & $13.49 \times 2.62$ & 6 & ATCA/6A \\
2013-07-21-23.25833\tablefootmark{f}\tablefootmark{g} & 655.63213 $\pm$ \phantom{1}3.63299 & \phantom{1}\phantom{1}2.8 & \phantom{1}\phantom{$-$}0.1300 $\pm$ \phantom{1}0.0200 & $11.63 \times 2.22$ & 6 & ATCA/6A \\
2016-09-06-08.50000 & 1798.01720 $\pm$ \phantom{1}0.08333 & \phantom{1}18.0 & \phantom{1}\phantom{$-$}0.0000 $\pm$ \phantom{1}0.0151 & $14.91 \times 14.58$ & 30 & ATCA/H168 \\
2016-09-07-08.75000 & 1799.02762 $\pm$ \phantom{1}0.07292 & \phantom{1}34.0 & \phantom{1}\phantom{$-$}0.0000 $\pm$ \phantom{1}0.0300 & $7.89 \times 7.84$ & 30 & ATCA/H168 \\
\hline
\end{tabular}
\tablefoot{ 
Proposal IDs and PI: ATCA: CX221, C2700, M.~Micha{\l}owski; EVN: RP018, M.~Micha{\l}owski;
\tablefoottext{a}{Mean time of the observation.}
\tablefoottext{b}{Time since the GRB explosion.}
\tablefoottext{c}{FWHM of the beam.}
\tablefoottext{d}{Position angle of the beam from North towards East.}
\tablefoottext{e}{For ATCA, PdBI, and VLA the array configuration is given.}
\tablefoottext{f}{Detection of the host galaxy.}
\tablefoottext{g}{The time span reflects the period over which the data was averaged, not the actual integration time.}
}
\end{table*}

\begin{table*}
\caption{Photometry of ESO580-49, the host galaxy of GRB 111005A. \label{tab:host}}
\centering
\begin{tabular}{lccll}
\hline\hline
$\lambda_{\rm obs}$ & Flux & Filter & Reference \\
(\micron)            & (mJy) &  & \\
\hline
0.1516 & $0.25 \pm 0.06$ & GALEXFUV & This paper \\
0.2267 & $0.530 \pm 0.037$ & GALEXNUV & This paper \\
0.3442 & $1.35 \pm 0.06$ & U & This paper \\
0.4390 & $3.30 \pm 0.29$ & B & \citet{lauberts89} \\
0.4770 & $3.272 \pm 0.006$ & B & This paper \\
0.6231 & $8.742 \pm 0.008$ & R & This paper \\
0.6390 & $6.9 \pm 0.6$ & R & \citet{lauberts89} \\
0.7625 & $10.404 \pm 0.019$ & I & This paper \\
0.7900 & $9.52 \pm 0.35$ & I & \citet{springob07} \\
1.25 & $15.3 \pm 0.7$ & J & \citet{2massmain} \\
1.64 & $17.3 \pm 1.0$ & H & \citet{2massmain} \\
2.17 & $14.9 \pm 1.3$ & K & \citet{2massmain} \\
3.6 & $9.078 \pm 0.017$ & IRAC1 & This paper \\
4.5 & $6.076 \pm 0.017$ & IRAC2 & This paper \\
60 & $347 \pm 43$ & IRAS60 & \citet{iras} \\
90 & $561 \pm 77$ & AKARI90 & \citet{akari} \\
100 & $957 \pm 252$ & IRAS100 & \citet{iras} \\
140 & $2016 \pm 264$ & AKARI140 & \citet{akari} \\
106310 & $0.124 \pm 0.016$ & 2.8 GHz & \citet{michalowski15} \\
126490 & $0.160 \pm 0.016$ & 2.35 GHz & \citet{michalowski15} \\
160320 & $0.192 \pm 0.018$ & 1.87 GHz & \citet{michalowski15} \\
215680 & $0.245 \pm 0.030$ & 1.39 GHz & \citet{michalowski15} \\
\hline
12 & $<140$ & IRAS12 & \citet{iras} \\
25 & $<146$ & IRAS25 & \citet{iras} \\
65 & $<252$ & AKARI65 & \citet{akari} \\
160 & $<1427$ & AKARI160 & \citet{akari} \\
870 & $<40$ & LABOCA870 & This paper \\
8817 & $<1.8$ & 34 GHz & This paper \\
16655 & $<1.5$ & 18 GHz & This paper \\
\hline
\end{tabular}
\tablefoot{ 
Upper limits are $2\sigma$. The archival data were compiled  from the NASA/IPAC Extragalactic Database with the appropriate reference shown in the last column. Radio limits are from our deepest afteglow photometry excluding the data in configurations with too high resolution, which resolves out the host extended emission.
}
\end{table*}

\begin{figure}
\begin{center}
\includegraphics[width= 0.5\propwidth]{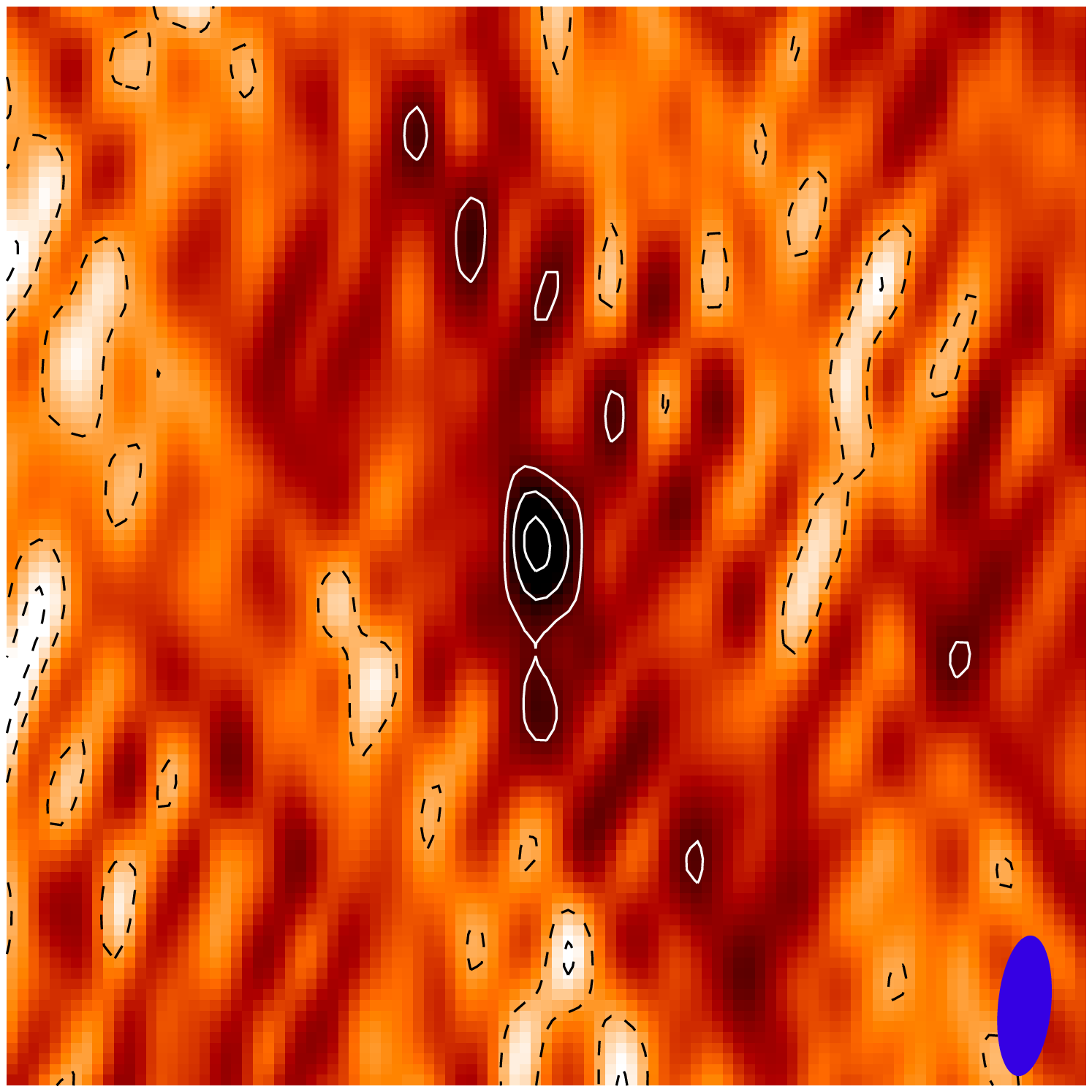}
\end{center}
\caption{VLBA image of the {\grb} afterglow on 2011 Oct 21 (16.5 days after the burst). North is up and east is to the left. The panel is  $0.01\arcsec\times0.01\arcsec$  ($2.7\mbox{ pc}\times2.7\mbox{ pc}$). The positive and negative contours are shown as {\it solid} and {\it dashed lines}, respectively at $-2, -1, 2, 3, 4\sigma$ with the rms of $0.15\,$mJy beam$^{-1}$. The beam ($1.31\times0.491\,\mbox{mas}$ FWHM) is shown as a {\it blue ellipse} in the bottom-right corner.
}
\label{fig:vlba}
\end{figure}

\begin{figure}
\begin{center}
\includegraphics[width= 0.5\propwidth]{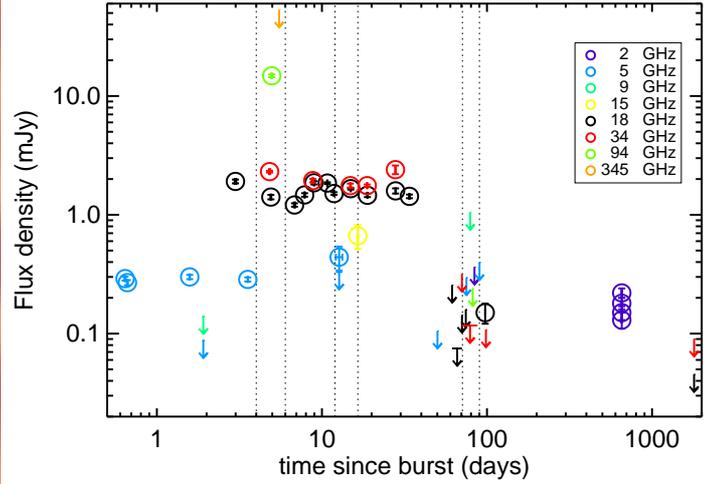}
\end{center}
\caption{Radio lightcurve of the afterglow of {\grb}. Datapoints are colour-coded by frequency. {\it Dotted lines} show the time intervals at which the spectral energy distributions are shown in Fig.~\ref{fig:sed}. The fluxes at $\sim650$ days are host detections (see Table~\ref{tab:after}). 
}
\label{fig:lightcurve}
\end{figure}

\begin{figure}
\begin{center}
\includegraphics[width= 0.5\propwidth]{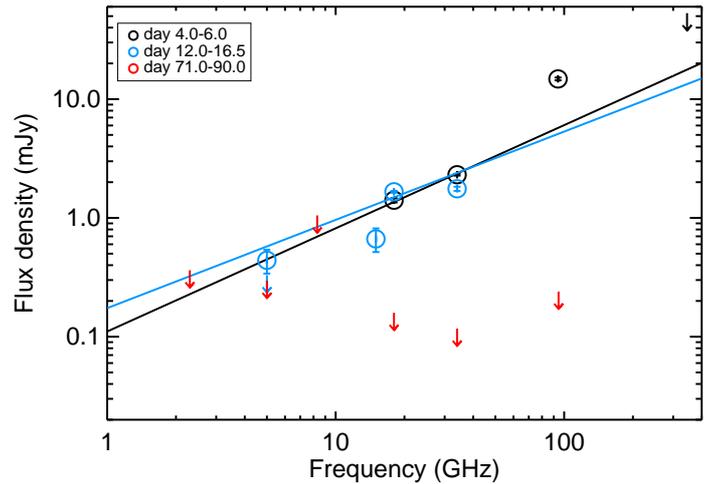}
\end{center}
\caption{Spectral energy distribution of the afterglow of {\grb}. Datapoints are colour-coded by the time at which they were obtained. 
The lines corresponds to a power-law fits (consistent with each other within errors) described in eq.~(\ref{eq:sed1}) and (\ref{eq:sed2}).
}
\label{fig:sed}
\end{figure}

\begin{figure}
\begin{center}
\includegraphics[width= 0.5\propwidth]{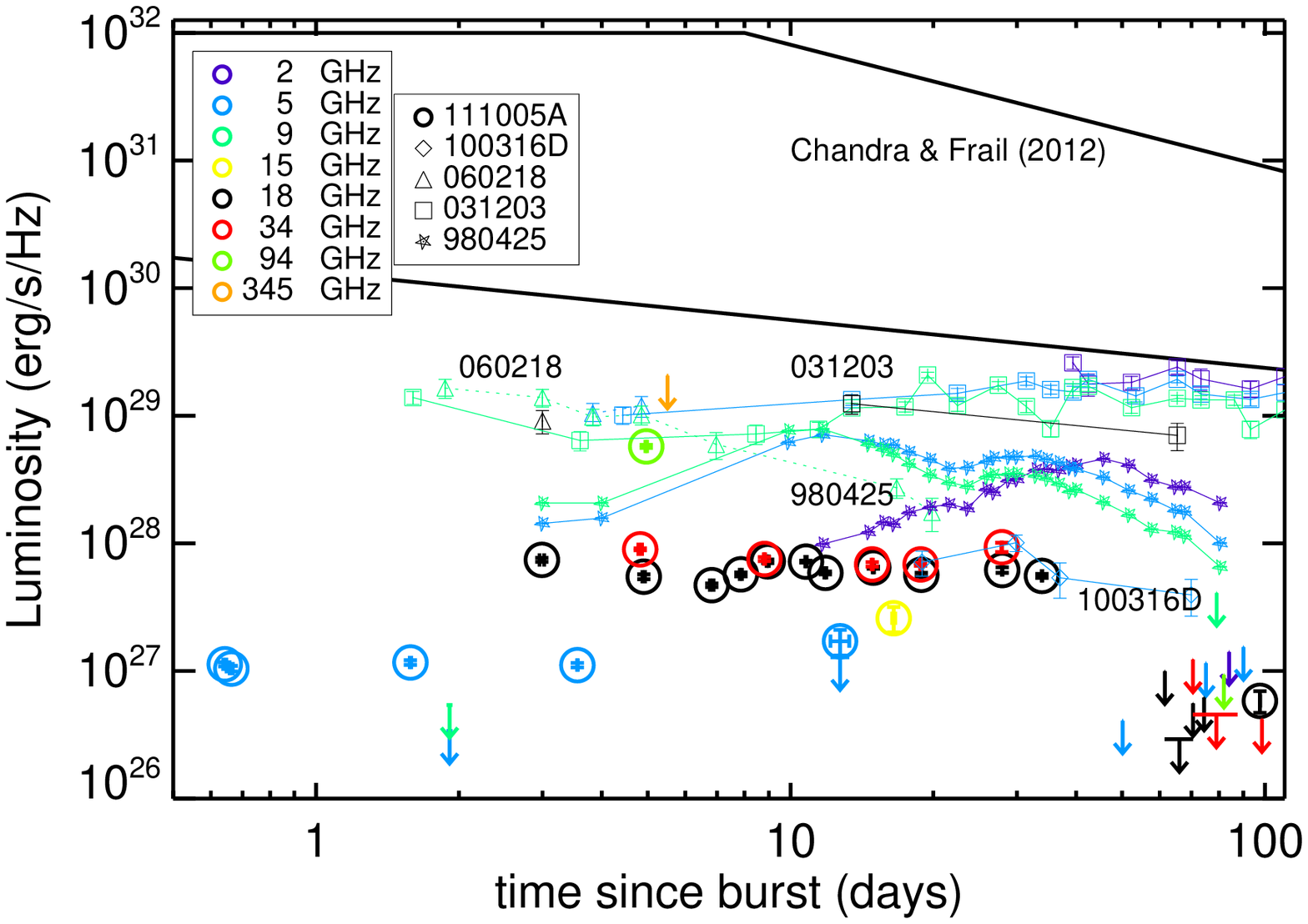}
\end{center}
\caption{Radio luminosity of the afterglow of {\grb} ({\it circles}), compared with cosmological GRBs (region marked by {\it black lines}; \citealt{chandra12}), GRB\,980425 \citep[{\it small stars};][]{kulkarni98}, GRB\,031203 \citep[{\it small squares};][]{soderberg04},  GRB\,060218 \citep[{\it small triangles};][]{soderberg06nature}, and GRB\,100316D \citep[{\it small diamonds};][]{margutti13}. Datapoints are colour-coded by frequency. 
}
\label{fig:lumt}
\end{figure}

\subsection{Radio}

We have obtained the data with the Australia Telescope Compact Array (ATCA) using the Compact Array Broad-band Backend \citep[CABB;][]{cabb} at 2--2\,000 days after the GRB event, i.e.~during 2011 Oct 7 to 2016 Sep 07,
detecting the afterglow up to a month after the event.  The array was in  various configurations during this period (see Table~\ref{tab:after}). The data reduction and analysis were done using the {\sc Miriad} package \citep{miriad,miriad2}. We have added the data obtained two years after the burst 
presented in \citet{michalowski15hi}.

We also observed {\grb}  at 5 GHz with the European VLBI Network
(EVN) 
during the 2011 Oct 17-18  realtime \mbox{e-VLBI} run in two parts,
between 11:23--13:17~UT on 17 Oct and between 13:08--15:08~UT
on 18 Oct. The participating telescopes were Effelsberg (Germany),
Jodrell Bank Mk2 (United Kingdom), Medicina (Italy), Onsala
(Sweden), Toru\'n (Poland), Yebes (Spain) and the phased-array
Westerbork Synthesis Radio Telescope (WSRT, Netherlands).
The field was centred at the position $\alpha=\mbox{}$14:53:07.78, $\delta=-$19:44:12.2
\citep{zauderer11gcn}. The target
was phase-referenced to the compact VLBI calibrator J1459-1810
at an angular distance of 2.2 degrees. Two candidate secondary
calibrators/check sources were selected from the Karl G.~Jansky Very Large Array
(VLA) NVSS survey \citep{nvss}\footnote{\url{http://www.cv.nrao.edu/nvss/}},
NVSS\,J145203.58-19438.00 (hereafter VLA1) and NVSS\,J145024.98-190915.2
(VLA2) at a distance of 15 and 51 arcmin, respectively. The phase-referencing cycle was 1m -- 1.5m -- 1.5m on J1459-1810, {\grb}
and VLA1, respectively, with every second cycle including VLA2 for
1.5 minutes. The second epoch was observed (also at 5~GHz) on
2011 Nov 24 between 8:25--12:35~UT with the same array with the
addition of Hartbeesthoek (South Africa). Since the angular distance
between our field and the Sun has decreased to slightly below 15 degrees,
we decided to use much closer VLA1 (detected during the first epoch)
as phase-reference source; we observed VLA1 for 1m20s and the target
for 2m30s per cycle.

The data were analysed with the NRAO Astronomical Image Processing
System\footnote{\urltt{http://www.aips.nrao.edu/cook.html}}  (AIPS; \citealt{aips}) using standard procedures as
described in the EVN Data Analysis Guide\footnote{\urltt{www.evlbi.org/user\_guide/guide/userguide.html}}.
and the maps were made in {\sc Difmap}  \citep{difmap}. {\grb}
was not detected at any of the epochs. On 2011 Oct 17-18  we achieved
a relatively high image noise of 100~$\mu$Jy~beam$^{-1}$ due to various
failures during the experiment, therefore we can give a 5$\sigma$
upper limit of 500~$\mu$Jy. On 24 November we achieved an image noise
of 35~$\mu$Jy~beam$^{-1}$ and the 5$\sigma$ upper limit is 175~$\mu$Jy.
We note however that the combination of small Sun-distance and low
declination of the target might have resulted in significant
correlation losses at this epoch.

We also observed {\grb} at 15.3 GHz by the Very Long Baseline Array (VLBA)
on 2011 Oct 21. 
The experiment lasted six hours and used a recording data rate of 512 Mbps (8 BBCs, dual sideband, 16 MHz filter, and 1-bit quantisation).  To further remove the residual tropospheric delay after the traditional phase-referencing calibration, two short (30 min) geodetic observations were scheduled at the beginning and end of the observations \citep{mioduszewski09}.  The source J1459$-$1810 was again observed as the main reference source. The cycle time was about 90\,s (30\,s on the calibrator, 50\,s on the {\grb}  or VLA1, $\sim10\,$s on slewing telescopes).  The nearby source VLA1 was also observed as a phase-referencing checker.  The total on-source time was 122 minutes on {\grb}  and 18 minutes on VLA1. The bright calibrator 1329-049 was observed as a fringe finder for a scan of 4 minutes.

The data were correlated by the software correlator DiFX \citep{difx2} with a frequency resolution 125 kHz (128 frequency points per subband) and an integration time of 1 second. Following the steps suggested by \citet{mioduszewski09}, we solved and applied the tropospheric delay. The rest of the steps are the same as for the EVN data reduction. 
The target 
was clearly detected in the image after all the calibration solutions were transferred from the calibrator to the targets. 

We have also obtained the radio observations with the WSRT.
Additionally, we have analysed the WSRT data alone taken during our EVN run. Data reduction and analysis were done using the  AIPS package. Only the early WSRT observations during the EVN run resulted in a detection.

\subsection{(Sub)mm}

We observed {\grb} with the Plateau de Bure Interferometer (PdBI)
in the compact `D' configuration on 2011 Dec 26  with the full array of six antennae in dual polarisation mode, and under excellent atmospheric weather conditions. The total observing time was 1.4~hr.  
The receivers were tuned to 94.5~GHz and the spectral bandwidth of the WideX correlator was 3.6~GHz. The flux calibration was done on MWC349 with a flux accuracy of 5\% and the data were reduced with the GILDAS software package CLIC and MAP. The FWHM of the beam is  $12\farcs2$$\times$$4\farcs6$  at PA$=$156.7~deg. The source was not detected.

We also performed submm ($870\,\mu$m) observations  on 2011  Oct 10, i.e.~five days after the burst 
using the Large  Apex BOlometer CAmera \citep[LABOCA;][]{laboca} mounted at the Atacama Pathfinder Experiment \citep[APEX;][]{apex}. A total of 0.9 hr of on-source data were obtained in the  on-off photometric mode. The weather was extremely poor with $2$--$3$ mm of precipitable water vapour, resulting in elevated noise and a non-detection.

\subsection{Optical and mid-IR}

\subsubsection{Imaging}


\begin{table*}
\caption{Log of optical/IR observations of {\grb}. In cases where it is relevant a limiting magnitude for the afterglow or supernova is shown.}
\begin{center}
\begin{tabular}{llllllll}
\hline
Date & MJD &  $\Delta T$ (days) & Telescope/Inst & Filter & exptime (s)  & Host (AB) &  OT limit(AB)\\
\hline
2011-10-05 & 55839.9773	& 0.64 & 		VLT/HAWK-I 		& 	K & 	540 &-  & $>21.4$ \\
2011-10-06 & 55840.9735	& 1.64 &		VLT/HAWK-I		& 	K  & 	600 & 14.084 $\pm$ 0.002  & - \\
2011-10-05 & 55839.9842	& 0.65 &		VLT/X-shooter acq & 	I  & 	60  & -  & $>$19.5 \\
2013-04-01 & 56383.1658	& 543.83 & 	VLT/X-shooter acq & 	R  & 120 & - & - \\
2011-10-14 & 55848.4856		& 9.15	& 	{\em Spitzer}/IRAC & 			3.6 &1140 & 14.005 $\pm$ 0.002 & 	$>21.9$  \\
2011-10-14 & 55848.4856		& 9.15	& 	{\em Spitzer}/IRAC & 			4.5 & 1140 & 14.441 $\pm$ 0.003 &    $>21.8$ \\
2012-04-14 & 56031.9333		& 192.60 &	{\em Spitzer}/IRAC & 			3.6 & 480 & - & -  \\
2012-04-14 & 56031.9333		& 192.60  &	{\em Spitzer}/IRAC & 			4.5 & 480 & -  & - \\
2012-05-21 & 56068.8483 	& 229.51 &	WHT/ACAM & 			g & 	400 & 15.113 $\pm$ 0.002 & -  \\  
2012-05-21 & 56068.9172		& 229.58 &	WHT/ACAM & 			r & 	400 & 14.046 $\pm$ 0.001 & -\\
2012-05-21 & 56068.9238  		& 229.59 &	WHT/ACAM & 			i & 	300 & 13.857 $\pm$ 0.002 & -  \\
\hline
\end{tabular}
\tablefoot{Proposal IDs and PI: VLT: 288.D-5004, N.~Tanvir and 088.D-0523, A.~Levan; WHT: W/2011B/21, A.~Levan; {\em Spitzer}: 80234,  PI: A.~Levan}
\end{center}
\label{optical_obs}
\end{table*}

Despite a location close to the Sun we obtained early multi-wavelength imaging observations of {\grb} utilising the VLT.
with additional later observations from the William Herschel Telescope (WHT)
A full log of observations is shown in Table~\ref{optical_obs}. Early observations were obtained with the X-shooter acquisition camera and the HAWK-I instrument at the VLT, taking place approximately 15 hours after the burst. Comparison observations for HAWK-I were obtained the following night (39 hours after burst), but further optical imaging was not obtained until 2012 May 21 with the WHT, and 2013 Apr 01, again with the X-shooter acquisition camera. 

The orbit of the {\em Spitzer} Space Telescope is such that it suffers from different periods of sun-block compared with ground based or low-Earth orbiting satellites. Because of this, {\em Spitzer} was able to obtain observations on 2011 Oct 14, 9 days after the burst, with a second comparison epoch obtained at 2012 Apr 14.
The first epoch was close to the expected time of the optical peak of any SNe associated with the burst, although an earlier peak is expected at longer wavelengths. 

Digital image subtraction images with the ISIS-II code of \citet{alard98} reveals no residuals in any of our images. Limiting magnitudes were estimated based on placing artificial sources within the images (in particular at the VLBA location) and then measuring the residual flux in the subtracted images. The location of the GRB is within the disc of the galaxy, although somewhat away from the nucleus and regions of highest surface brightness (see Fig.~\ref{fig:im}), such that the photon noise from the galaxy have a smaller impact on the limiting magnitudes than might otherwise be the case, although this contribution is larger in the {\em Spitzer} observations which have a rather poor PSF. 

In several cases subtractions offer limited potential. In the K-band there is only a short baseline (1 day) between each epoch of observations. While this is likely very sensitive for GRB afterglows (it is a factor of 2.5 in time) it is less so for any SNe, which may only vary by a few hundredths to a few tenths of a magnitude in this time frame in the K-band (although the early light curves of SNe may also show a strong rise). Hence, the limit is not necessarily reflecting a limit from a ``transient'' free observations. We note that insertion of manual point sources in this case would result in a clear detection of a point source superimposed on the stellar field of the galaxy for sources brighter than $K \sim 18$\,mag. Similarly, for our X-shooter observations there exists no later time observations taken with the same instrument, while the lack of calibrations taken with these data (taken with an acquisition camera) means also complicates subtractions. These limits are therefore obtained by both noting the point at which a point source becomes clearly visible in the images when inserted (necessarily a qualitative judgement) and by subtraction of the images from later observations taken both with the same camera, but a different filter, and in the same filter, but with a different camera. In practice the limitations of both approaches yield rather similar answers in each case.

Photometric calibration was performed relative to the AAVSO Photometric All-Sky Survey (APASS; \citealt{apass}) for our optical observations and the Two Micron All Sky Survey (2MASS; \citealt{2mass,2massmain}) in the infrared (IR), known zero points were used for {\em Spitzer} observations. In the case of our X-shooter observations a limited number of APASS secondary standards were visible in our FOV (none in the first epoch i-band observations). Therefore we initially calibrate the WHT observations to APASS, and then to the X-shooter acquisition data. We therefore calibrate all of our optical data to SDSS filters. Astrometric calibration was performed relative to the Third US Naval Observatory CCD Astrograph Catalog (UCAC3; \citealt{ucac3}), and yielded a WCS fit to better than 0.1\arcsec\ in most cases. This enables the VLBA position to be placed on our images to sub-pixel accuracy. 

We obtained a single orbit of observations with the Hubble Space Telescope (HST), utilising Wide Field Camera 3 on 2015 Jun 03 (proposal 13949, PI: A.~Levan). Observations were obtained in the F438W and F606W with the UVIS channel, and in the F160W filter in the IR channel. The exposure times for each filter were 1044, 686 and 306 respectively, and the data were reduced via {\tt astrodrizzle} in the standard fashion. To precisely place the location of {\grb} on these images we subsequently align them to 2MASS observations using 6 stars in the field (one is omitted because of a significant offset from its 2MASS position, likely due to high proper motion). The result RMS of the fit to the world co-ordinate system is $\sim 0.1\arcsec$ in each axis. 

\subsubsection{Spectroscopy}

In addition to our imaging observations we obtained spectroscopy with the WHT on 2012 May 21, using the ISIS spectrograph with the R600B and R600R gratings. A total of $4 \times 300$ s exposures were obtained in each arm, with the slit aligned to run through the major axis of the galaxy. The slit position is shown in Fig.~\ref{fig:slit}.

The host galaxy of {\grb} was also observed with the X-shooter spectrograph mounted on the ESO VLT on 2013 Apr 01 (proposal 090.A-0088, PI: J.~Fynbo). The observation consisted of 2600 sec exposures at a fixed slit position of 151 degrees East of North illustrated in Fig.~\ref{fig:Xslit}. This slit position covers both the centre of the galaxy and the radio position of the GRB.

\subsection{Archival data}

We have analysed the VLA data taken during 2011 Nov 05 -- Oct 14  (%
\citealt{zauderer11gcn}). VLA was in the D configuration.
We first applied the VLA pipeline\footnote{\url{https://science.nrao.edu/facilities/vla/data-processing/pipeline}} written in the Common Astronomy Software Applications ({\sc Casa}) package \citep{casa}. Then the data were imaged and analysed in {\sc Casa}. 

We have obtained the archival {\it Swift}/UVOT \citep{uvot} $u$-band data taken on 2011 Dec 29 
(85 days after the burst) and 2012 Jan 04 
(91 days after the burst). We combined all images from a given epoch and subtracted the later combined image from the earlier one, which did not reveal any variable source (Fig.~\ref{fig:imu}). Therefore we averaged all the data for the host galaxy analysis. We measured its flux in a $67.5\arcsec$ aperture.

We obtained the host galaxy photometry from the NASA/IPAC Extragalactic Database, including IRAS and AKARI data. We also added our radio data from \citet{michalowski15hi}.  
Finally, we have measured its ultraviolet (UV) emission from the {\it GALEX} \citep{galex1,galex2}\footnote{{\it Galaxy Evolution Explorer}; \url{http://galex.stsci.edu/}} archive,
using $67.5\arcsec$ apertures.

\section{Methods}
\label{sec:method}
\label{sec:methhost}

For the host galaxy emission we applied the SED fitting method detailed in \citet[][see therein a discussion of the derivation of galaxy properties and typical uncertainties]{michalowski08,michalowski09,michalowski10smg,michalowski10smg4,michalowski12mass,michalowski14mass} which is based on 35\,000 templates from the library of \citet{iglesias07} plus some templates of \citet{silva98} and \citet{michalowski08}, all of which were developed using {\grasil}\footnote{\url{adlibitum.oats.inaf.it/silva/grasil/grasil.html}} \citep{silva98}. They are based on numerical calculations of radiative transfer within a galaxy, which is assumed to be a triaxial axisymmetric system with diffuse dust and dense molecular clouds, in which stars are born.

The templates cover a broad range of galaxy properties from quiescent to starburst and span an $A_V$ range from $0$ to $5.5$ mag. The extinction curve \citep[fig.~3 of][]{silva98} is derived from the modified dust grain size distribution of \citet{draine84}.
The star formation histories are assumed to be a smooth Schmidt-type law \citep[i.e., the SFR is proportional to the gas mass to some power; see][for details]{silva98} with a starburst (if any) on top of that, starting $50$ Myr before the time at which the SED is computed. There are seven free parameters in the library of \citet{iglesias07}: the normalisation of the Schmidt-type law, the timescale of the mass infall, the intensity of the starburst, the timescale for molecular cloud destruction, the optical depth of the molecular clouds, the age of the galaxy and the inclination of the disk with respect to the observer.

We also used {\magphys}\footnote{\url{www.iap.fr/magphys}} \citep[Multi-wavelength Analysis of Galaxy Physical Properties;][]{dacunha08}, which is an empirical, physically-motived SED modelling code that is based on the energy balance between the energy absorbed by dust and that re-emitted in the infrared. We used the \citet{bruzualcharlot03} stellar population models and adopted the \citet{chabrier03} IMF. 

Similarly to {\grasil}, in {\magphys}, two dust media are assumed: a diffuse interstellar medium (ISM) and dense stellar birth clouds. Four dust components are taken into account: cold dust ($15$--$25$ K), warm dust ($30$-$60$ K), hot dust ($130$--$250$ K) and polycyclic aromatic hydrocarbons (PAHs). A simple power-law attenuation law is assumed.

We excluded some data from the SED modelling.
The IRAS $100\,\micron$ and  AKARI $140\,\micron$ fluxes are likely affected by poor resolution and are overestimated (like in the case of GRB 980425 host with the $160\,\micron$ {\it Spitzer} fluxes a factor of two higher than the {\it Herschel}/PACS fluxes, compare \citealt{lefloch12} and \citealt{michalowski14}).  On the other hand, the  ATCA radio observations from \citet{michalowski15hi} resolved the host (beamsize from $\sim30\arcsec\times4\arcsec$ to $10\arcsec\times2\arcsec$), so the flux is likely underestimated.

\section{Results}
\label{sec:res}

\renewcommand{\tabcolsep}{0.1cm}
\begin{table*}
\caption{{\sc Grasil} results from the SED fitting. \label{tab:grasilres}}
\scriptsize
\begin{center}
\begin{tabular}{cccccccccc}
\hline\hline
 $\log L_{\rm IR}$ & $\mbox{SFR}_{\rm IR}$ & $\mbox{SFR}_{\rm SED}$ & $\mbox{SFR}_{\rm UV}$ & $\mbox{sSFR}_{\rm SED}$ & $\log M_*$ & $\log M_{\rm dust}$ & $\log T_{\rm dust}$ & $A_V$ & $\log\mbox{age}_M$ \\
 ($L_\odot$) & ($M_\odot\,\mbox{yr}^{-1}$) & ($M_\odot\,\mbox{yr}^{-1}$) & ($M_\odot\,\mbox{yr}^{-1}$) & ($\mbox{Gyr}^{-1}$) & ($M_\odot$) & ($M_\odot$) & (K) & (mag) & (yr) \\
 (1) & (2) & (3) & (4) & (5) & (6) & (7) & (8) & (9) & (10) \\
\hline
 9.63 & 0.41 & 0.38 & 0.15 & 0.07 & 9.72 & 6.35 & 39 & 0.14 & 9.78 \\
\hline
\end{tabular}
\tablefoot{ (1) $8$--$1000\,\mu$m infrared luminosity. (2) star formation rate from $L_{\rm IR}$ \citep{kennicutt}. (3) star formation rate from SED modelling. (4) star formation rate from UV emission \citep{kennicutt}. (5) specific star formation rate ($\equiv\mbox{SFR}_{\rm SED}/M_*$). (6) stellar mass. (7) dust mass. (8) dust temperature. (9) mean dust attenuation at $V$-band. (10) mass-weighted age. 
}
\end{center}
\end{table*}

\renewcommand{\tabcolsep}{0.1cm}
\begin{table*}
\caption{{\sc Magphys} results from the SED fitting. \label{tab:magphysres}}
\scriptsize
\begin{center}
\begin{tabular}{ccccccccccccccc}
\hline\hline
$\log L_{\rm IR}$ & SFR & sSFR & $\log M_*$ & $\log M_d$ & $\tau_V$ & $T_{\rm cold}$ & $\xi_{\rm cold}$ & $T_{\rm warm}$ & $\xi_{\rm warm}$ & $\xi_{\rm hot}$ & $\xi_{\rm PAH}$ &  $f_\mu$ & $\log\mbox{age}_M$\\
($L_\odot$) & ($M_\odot\,\mbox{yr}^{-1}$) & (Gyr$^{-1}$) & ($M_\odot$) & ($M_\odot$) & & (K) & & (K) & & & & & (yr)\\
(1) & (2) & (3) & (4) & (5) & (6) & (7) & (8) & (9) & (10) & (11) & (12) & (13) & (14)\\
\hline
 $9.58^{+0.09}_{-0.08}$ & $0.42^{+0.06}_{-0.05}$ & $0.09^{+0.03}_{-0.02}$ & $9.68^{+0.13}_{-0.09}$ & $6.57^{+0.43}_{-0.40}$ & $0.77^{+0.86}_{-0.17}$ & $19.7^{+3.4}_{-2.8}$ & $0.29^{+0.10}_{-0.10}$ & $43^{+11}_{-8}$ & $0.49^{+0.13}_{-0.11}$ & $0.10^{+0.05}_{-0.05}$ & $0.10^{+0.06}_{-0.06}$ & $0.38^{+0.11}_{-0.12}$ & $9.94^{+0.10}_{-0.10}$ &  \\
\hline
\end{tabular}
\tablefoot{(1) $8-1000\,\mu$m infrared luminosity. (2) star formation rate from SED modelling. (3) specific star formation rate ($\equiv\mbox{SFR}/M_*$). (4) stellar mass. (5) dust mass. (6) average $V$-band optical depth ($A_V=1.086\tau_V$).  (7) temperature of the cold dust component. (8) contribution of the cold component to the infrared luminosity. (9) temperature of the warm dust component. (10) contribution of the warm component to the infrared luminosity. (11) contribution of the hot ($130$--$250$ K, mid-IR continuum) component to the infrared luminosity. (12) contribution of the PAH component to the infrared luminosity. (13) contribution of the ISM dust (as opposed to birth clouds) to the infrared luminosity. (14) mass-weighted age.}
\end{center}
\end{table*}

\begin{figure}
\begin{center}
\includegraphics[width= 0.5\propwidth]{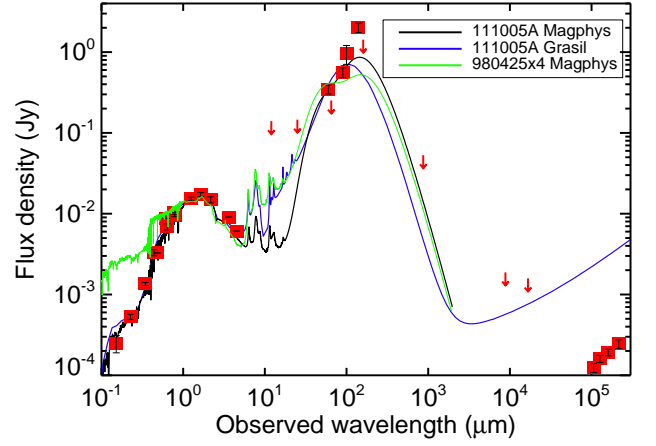}
\end{center}
\caption{Spectral energy distribution of ESO 580-49, the host galaxy of {\grb}. Datapoints are shown as {\it red squares and arrows}, whereas {\grasil} and {\magphys} models are shown as a {\it blue and black lines}, respectively. The data at $100$, $140\,\micron$ and in the radio were not used in the modelling due to either too poor spatial resolution, or resolving out the extended emission (see Sec.~\ref{sec:methhost}). The SED of the GRB 980425 host \citep{michalowski14} scaled up by a factor of 4 is shown for comparison ({\it green line}). 
}
\label{fig:sedhost}
\end{figure}

 Our best position of the {\grb} afterglow comes from the VLBA observations with $1.31\times0.491\,\mbox{mas}$ FWHM beam (Fig.~\ref{fig:vlba}). This results in the position of the radio afterglow of $\alpha=\mbox{14:53:07.8078276}$,   $\delta= -\mbox{19:44:11.995387}$ (J2000) with the $1\sigma$ error of $0.2\,$mas. The source is not resolved with the $3\sigma$ upper limit on the angular size of $< 0.38\,$mas.
  
 All data obtained during our multi-facility campaign are presented in Table~\ref{tab:after}, 
 whereas the host galaxy photometry is presented in Table~\ref{tab:host}.
 The lightcurve and the 3-epoch spectral energy distribution of the afterglow are shown in Fig.~\ref{fig:lightcurve} and \ref{fig:sed}, respectively. Fig.~\ref{fig:lumt} shows the luminosity of the radio afterglow compared with cosmological GRBs \citep{chandra12} and local low-luminosity GRB\,980425 \citep{kulkarni98}, GRB\,031203 \citep{soderberg04},  GRB\,060218 \citep{soderberg06nature}, and GRB\,100316D \citep{margutti13}.

 Fig.~\ref{fig:im} shows the images of the host galaxy at  the UV and {\it Spitzer} wavelengths.
 The spectral energy distribution of the host galaxy is shown in Fig.~\ref{fig:sedhost}.  
 The galaxy properties derived using {\grasil} and {\magphys} are shown in Tables~\ref{tab:grasilres} and \ref{tab:magphysres}, respectively. All results of these two codes are consistent,  especially the stellar mass estimates, which results from the good optical/near-infrared data coverage.
We note that {\grasil} uses the mass absorption coefficient $\kappa(1.2\,\rm{mm})=0.67\, \cmg$ \citep{silva98}, i.e.~$\kappa(850\,\micron)=1.34\,\cmg$ (assuming $\beta=2$), whereas
 {\magphys} uses the value $1.7$ times smaller $\kappa(850\,\micron)=0.77\,\cmg$ \citep{dacunha08,dunne00}, which should result in a higher dust mass. 
 Indeed,  {\magphys} predicts a factor of $1.7$ larger dust mass, so this difference can be fully explained by the difference in $\kappa$.

\begin{figure}
\begin{center}
\includegraphics[angle=0,width=0.5\textwidth,clip]{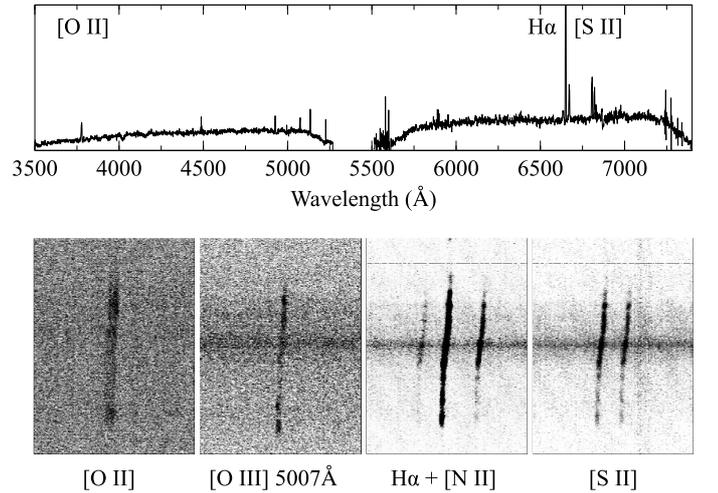}
\end{center}
\caption{{\it Top}: Spectrum of the {\grb} host added over the entire extend of the slit (Fig.~\ref{fig:slit}). Some emission lines are marked, and the {\it bottom} panels show their two-dimensional spectra. The horizontal axis corresponds to the wavelengths and the vertical axis to the position along the slit (400 pixels, i.e.~80\arcsec). The rotation curve is clearly visible with each line.  The drop at $\sim5300$\,{\AA} is due to the dichroic gap between the blue and red arms of the ISIS.}
\label{fig:spec}
\end{figure}

\begin{figure*}
\begin{center}
\begin{tabular}{cc}
\includegraphics[angle=0,height=0.4\textwidth,clip]{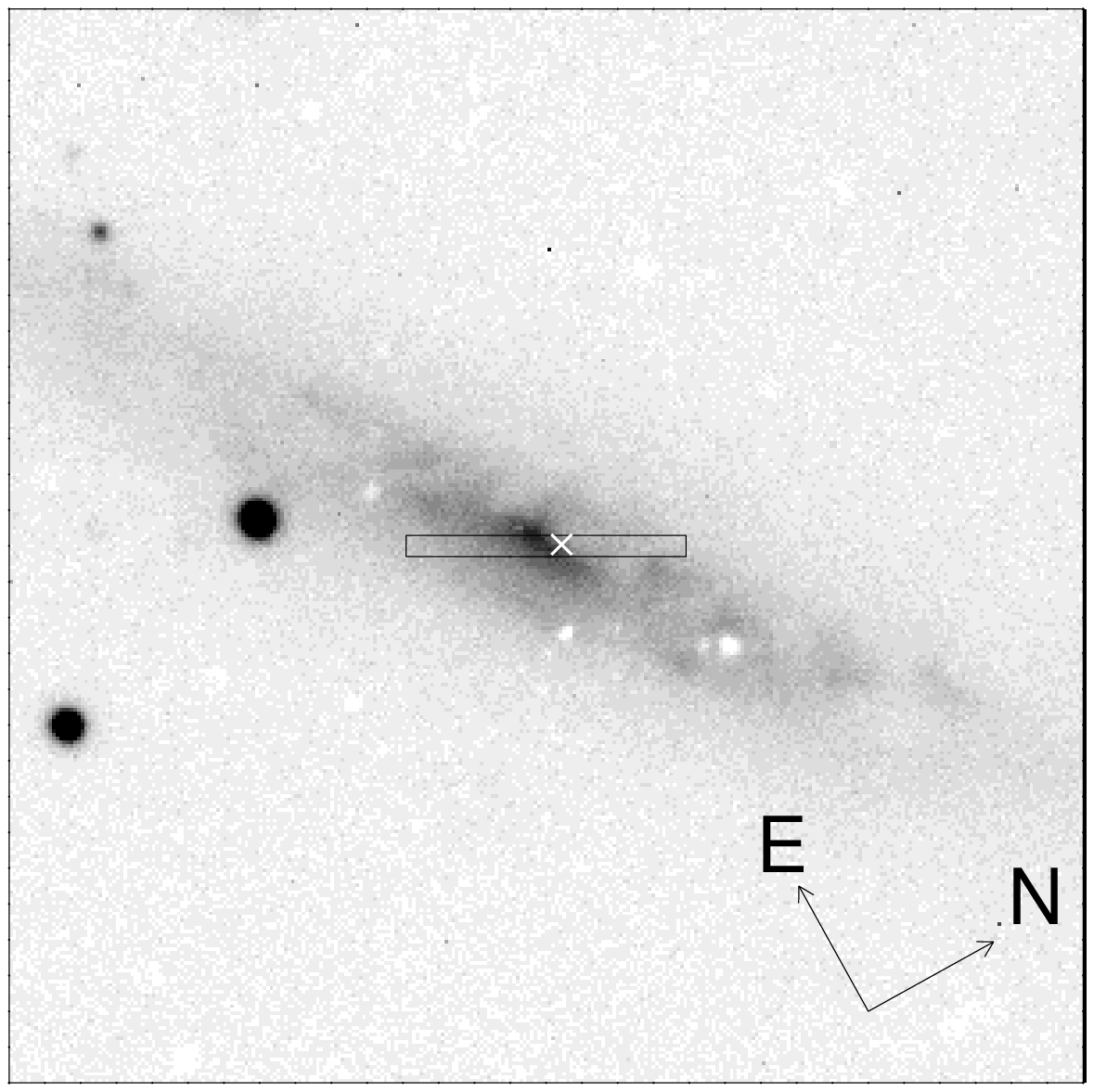} &
\includegraphics[angle=0, height =0.4\textwidth,clip]{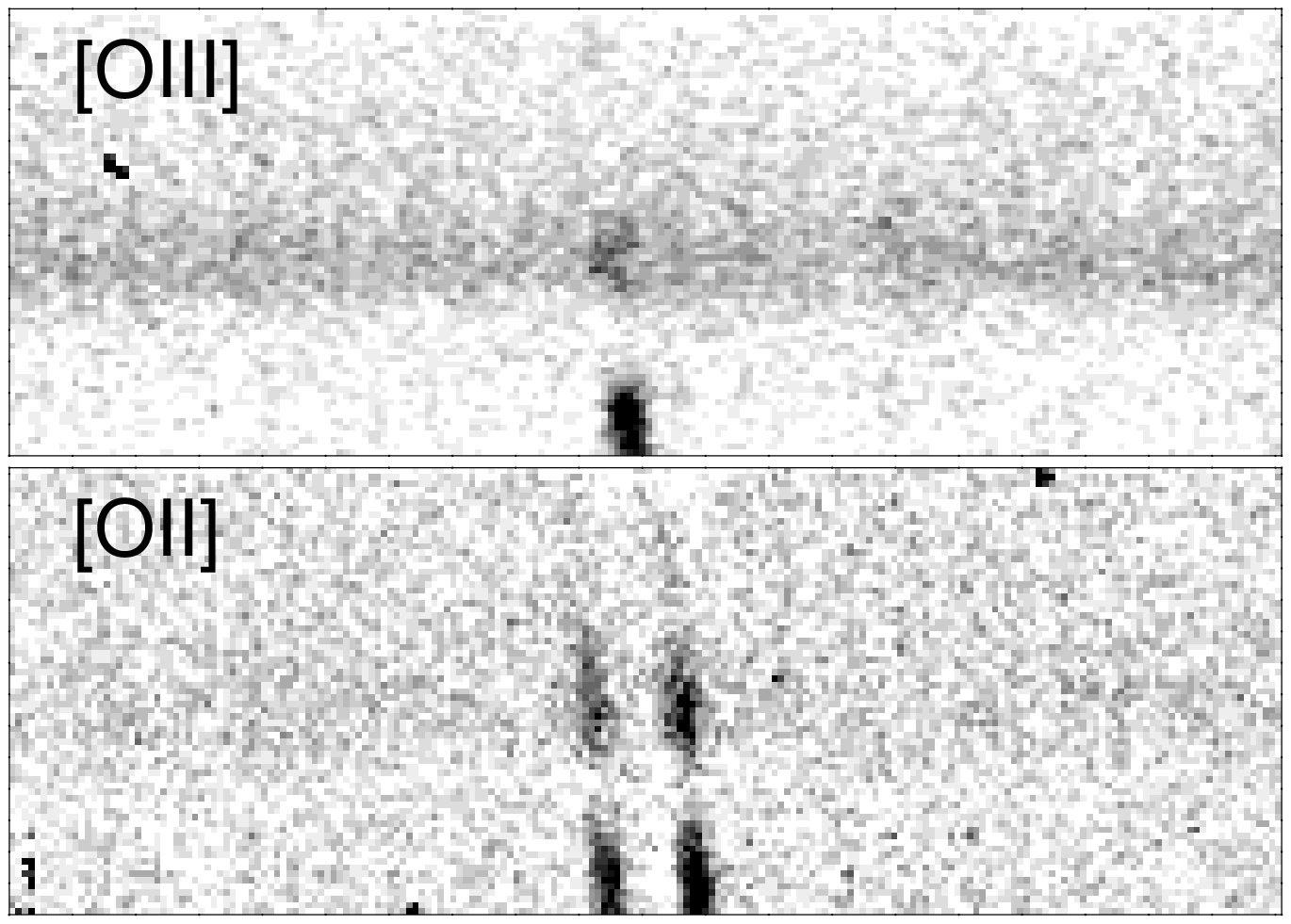} \\
\end{tabular}
\end{center}
\caption{{\it Left}:  orientation of the VLT/X-shooter slit. The {\it white cross} marks our VLBA position of {\grb}. {\it Right}: two-dimensional spectra. The horizontal axis corresponds to the wavelengths and the vertical axis to the position along the slit. The rotation curve is clearly visible with each line.  The emission to the left corresponds to the galaxy center, whereas the one to the right is offset $\sim4.5\arcsec$ to the Northwest and have a much harder  ionising flux as it exhibits much higher [\ion{O}{iii}]/[\ion{O}{ii}] ratio.
}
\label{fig:Xspec}
\label{fig:Xslit}
\end{figure*}

\begin{figure}
\begin{center}
\includegraphics[angle=0,width=0.5\textwidth,clip,viewport= 35 62 585 510]{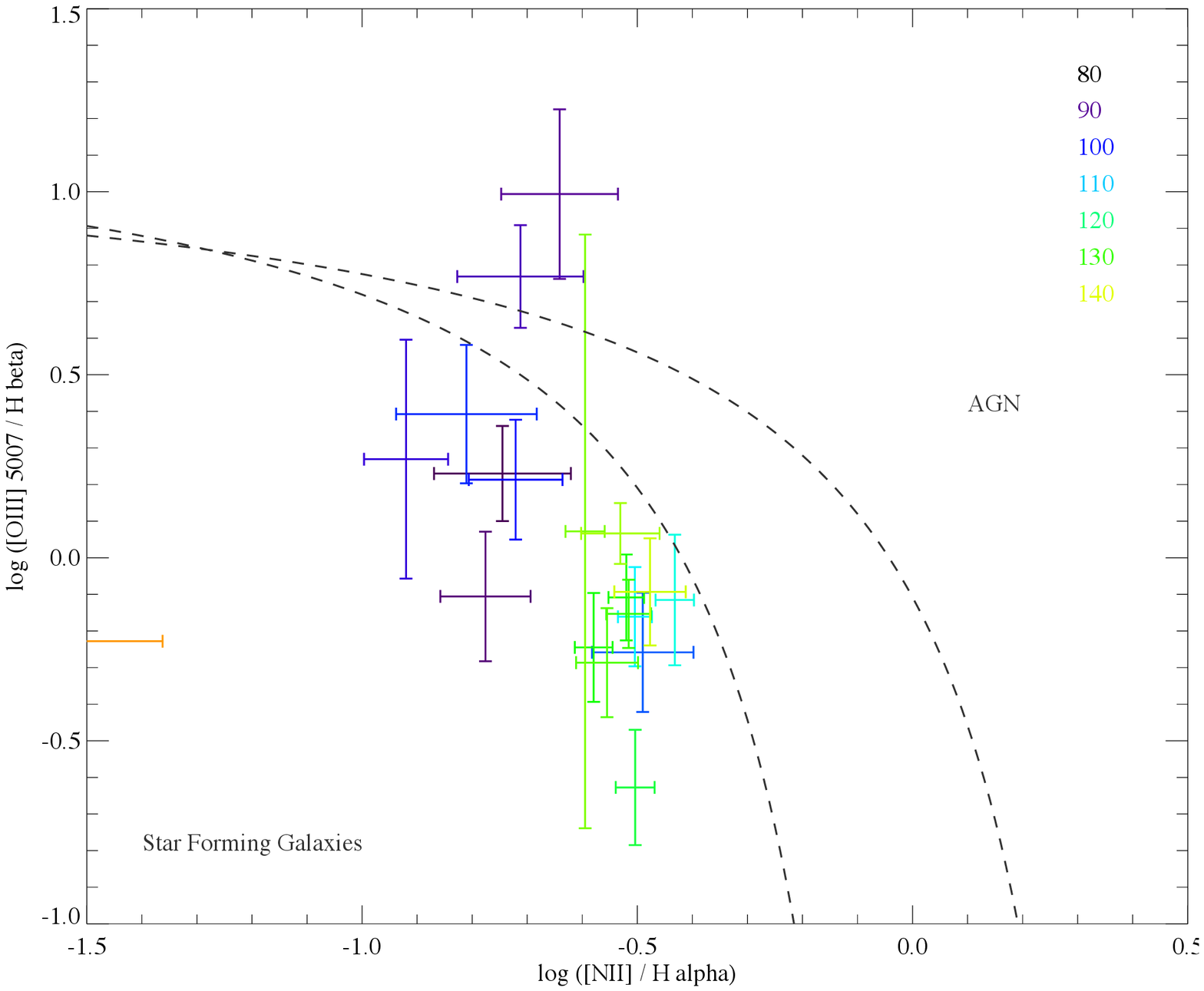}
\end{center}
\caption{BPT diagnostic \citep{bpt} along the slit of the host of {\grb}. Most of the regions are consistent with star forming activity. The nucleus lies at approximately 115\arcsec. The regions along the slit are defined in Fig.~\ref{fig:slit}.}
\label{fig:bpt}
\end{figure}

\begin{figure}
\begin{center}
\includegraphics[angle=0,width=0.5\textwidth,clip,viewport=35 60 585 510]{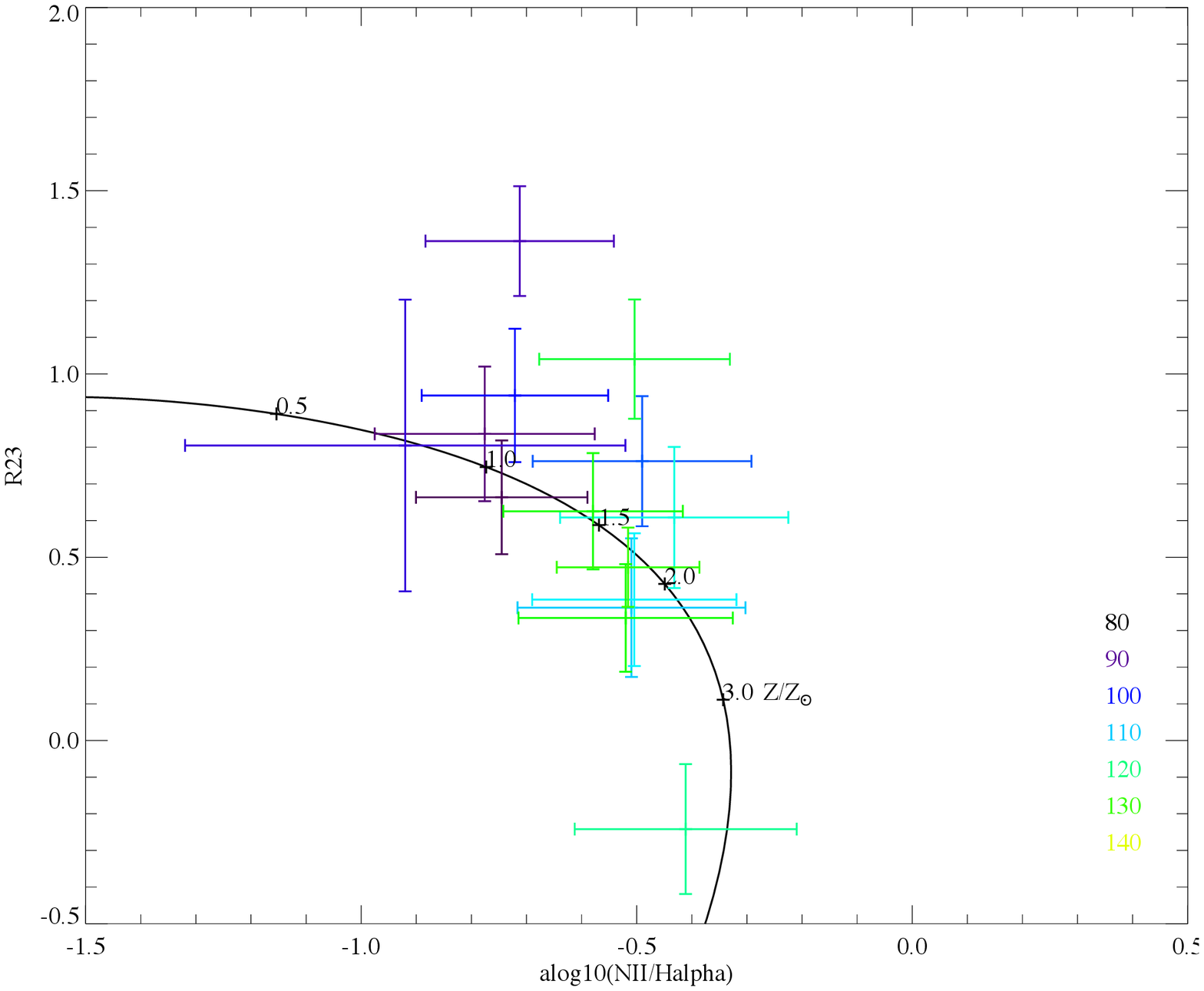}
\end{center}
\caption{The $R_{23}$ metallicity diagnostic along the host of {\grb}. The nucleus lies at approximately 115\arcsec.The regions along the slit are defined in Fig.~\ref{fig:slit}.}
\label{fig:metal}
\end{figure}

\begin{figure}
\begin{center}
\includegraphics[angle=90,width=0.5\textwidth,clip,viewport=10 165 562 710]{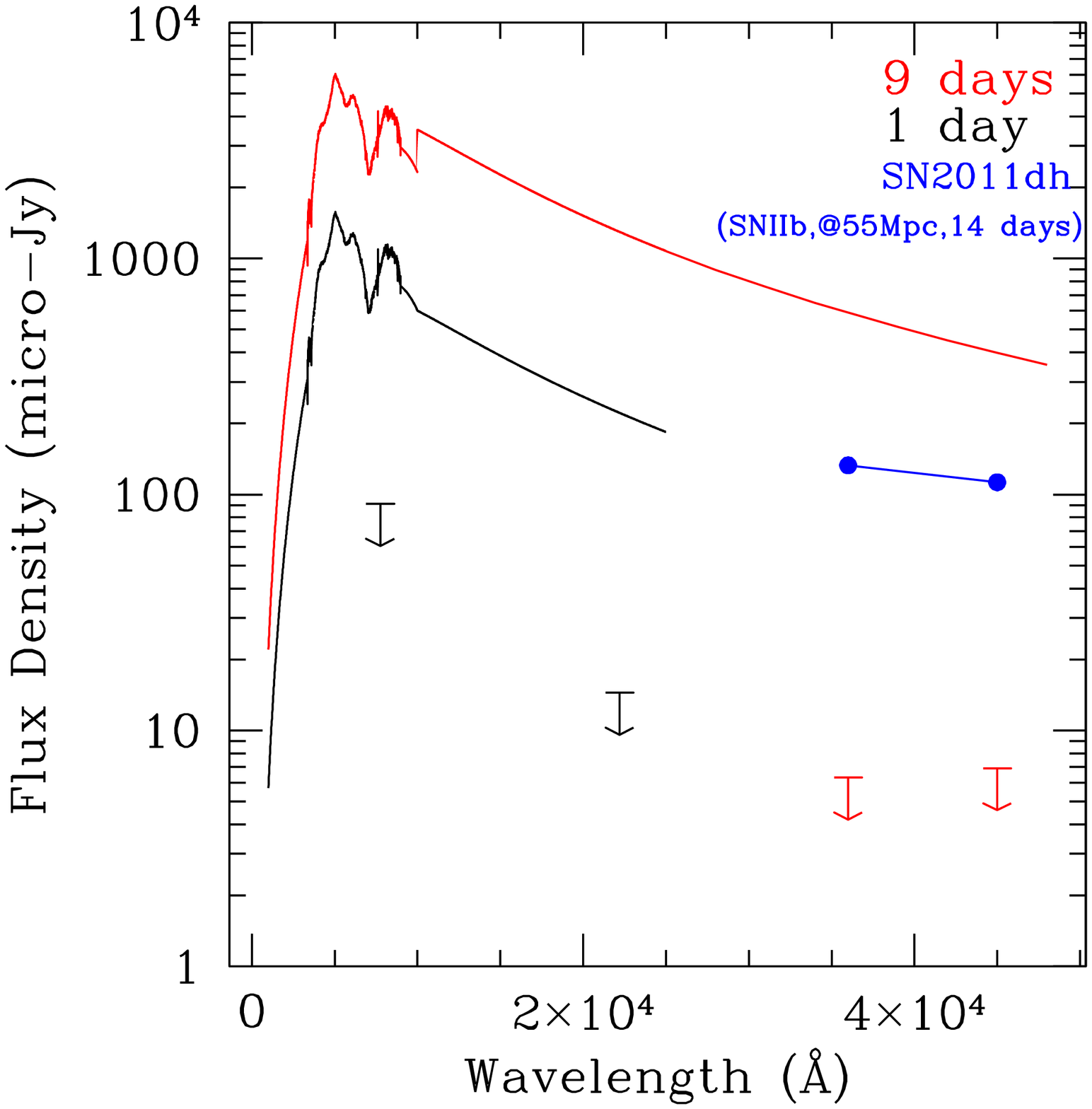}
\end{center}
\caption{Limits on supernova-like emission from GRB111005A (arrows). The lines show the models from \citet{levan05}, scaled to a distance of 55 Mpc. In the case of the 9-day epoch we have used a model from 7-days, and extended the template from 2.2 to 4.8\,{\micron} based on a purely thermal extrapolation. We also show the observations of the type IIb SNe 2011dh in M51 \citep{helou13}, as it would appear at 55 Mpc. }
\label{snlimits}
\end{figure}

The WHT spectrum (Fig.~\ref{fig:spec}) shows a wealth of starforming emission lines (including O[{\sc ii}], O[{\sc iii}], H$\beta$, H$\alpha$, N[{\sc ii}], S[{\sc ii}]) visible across the galaxy disc, as well as a
strong central bulge which appears as a near point source running through the spectra. The Baldwin-Phillips-Terlevich \citep[BPT;][]{bpt} diagnostics (Fig.~\ref{fig:bpt}) 
are generally consistent with star forming activity (not active galactic nuclei [AGN]), including that close to the nuclear regions of the galaxy, although we do note a region approximately 10\arcsec\ from the nucleus in which AGN-like ratios are observed. The metallicity around the GRB region and nucleus of the galaxy, as inferred by the $R_{23}$ diagnostic is relatively high (Fig.~\ref{fig:metal}), and suggests the GRB is born in a region with metallicity in excess of solar. 

\begin{figure}
\begin{center}
\includegraphics[angle=0,width=0.5\textwidth,clip,viewport=15 15 660 585]{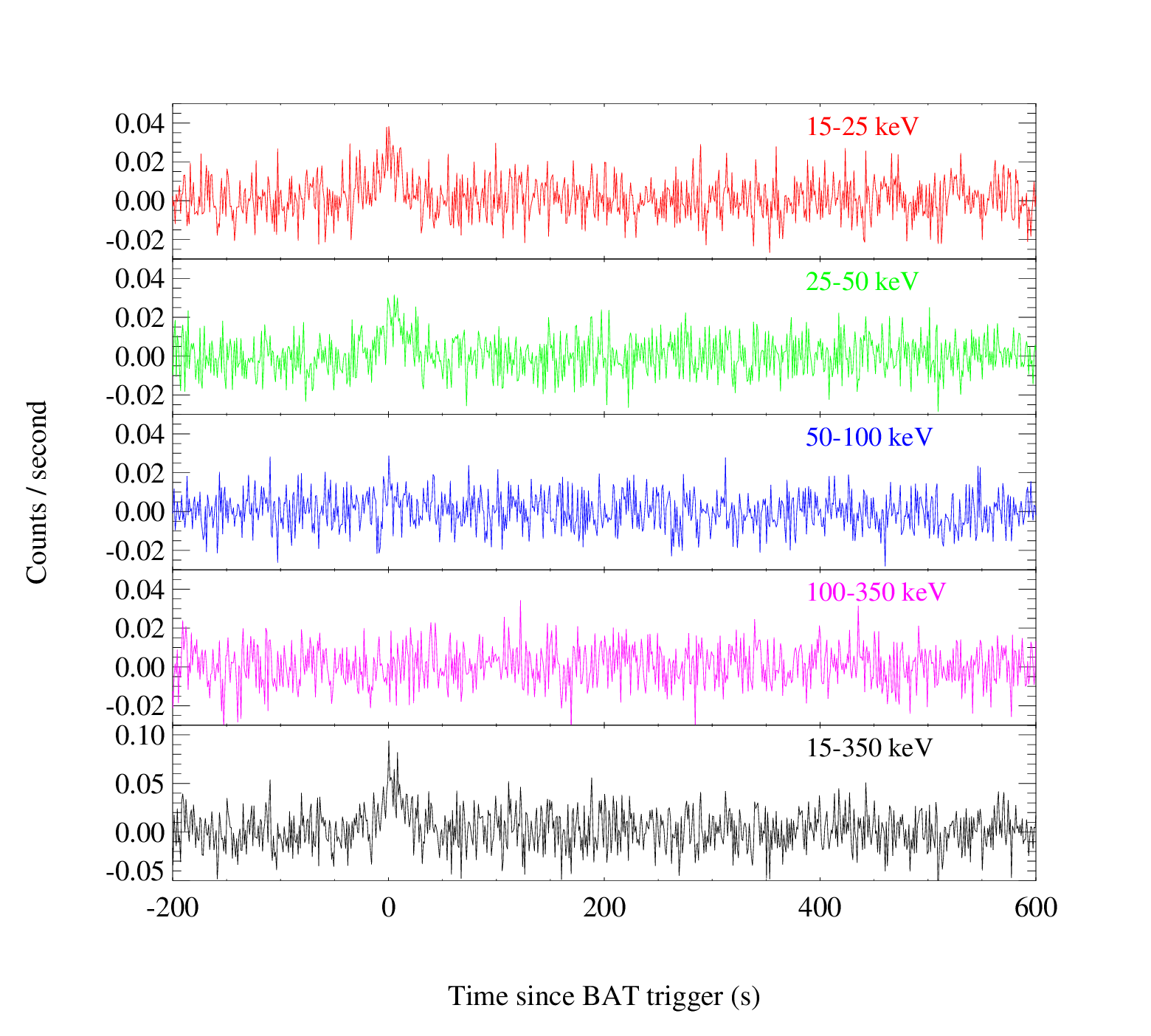}
\end{center}
\caption{Prompt emission lightcurve of {\grb}. The first four panels show four {\it Swift}/BAT energy bands, whereas the last one show the combined emission.}
\label{fig:prompt}
\end{figure}

None of our subtractions yield any obvious residual emission in the optical or near-infrared. At 55 Mpc we would have expected either a GRB afterglow or an associated SNe to be extremely bright. SN~1998bw would have appeared at a magnitude of $R < 16.5$ at the time of our first optical observations, and would have been clearly visible as a point source on the host galaxy. While these 
observations could be rendered of limited value by (unknown) extinction within the host galaxy, our {\em Spitzer} observations effectively remove that concern, and suggest that 
any supernova associated with {\grb} must have been a factor $>50$ fainter than SN~1998bw. In Fig.~\ref{snlimits} we plot these limits graphically compared with both
the expectations of models of broad lined SN Ic \citep[][in the case of the mid-IR extrapolated into the mid-IR using a simple blackbody model]{levan05}, as well observations of SN~2011dh, a SN IIb with good {\em Spitzer} observations \citep{helou13}. The Spitzer observations largely remove any concerns relating to extinction since even an $A_V=30$\,mag would yield $A_{3.6} = 1.5\,$mag, and so we would expect to have detected the resulting SNe. Indeed, we note that the deep IR limits ($\sim22$ mag AB) imply an absolute magnitude of $M_{IR} > -12$, comparable to the magnitude of the faint red transient detected by \citet{kulkarni07} in M85, which has been suggested to be a stellar merger. Hence we cover the full range of expected supernova properties, and probe into regions normally occupied by supernova impostors.

\begin{figure*}
\begin{center}
\includegraphics[angle=90,width=0.5\textwidth,clip]{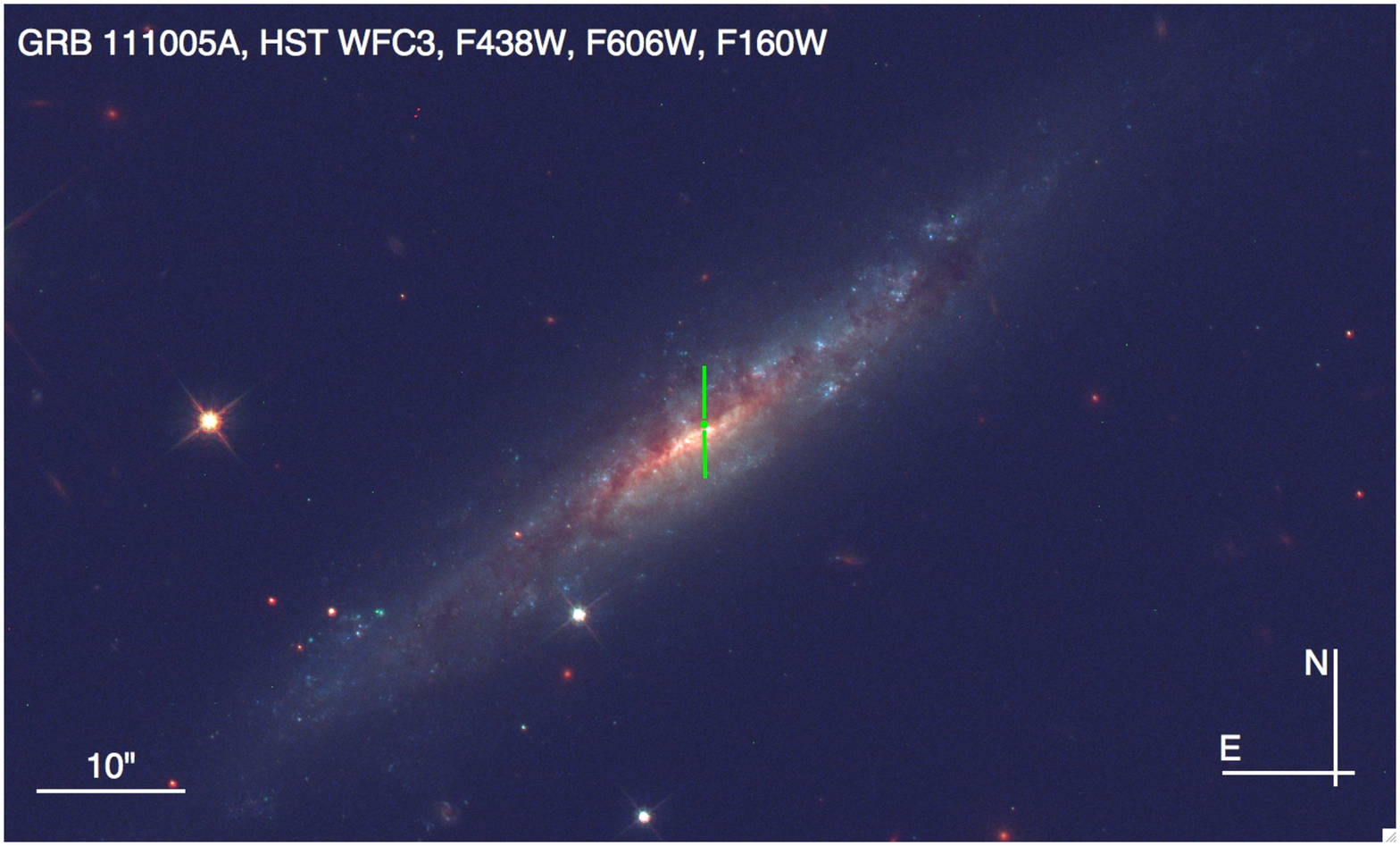}%
\includegraphics[angle=90,width=0.5\textwidth,clip]{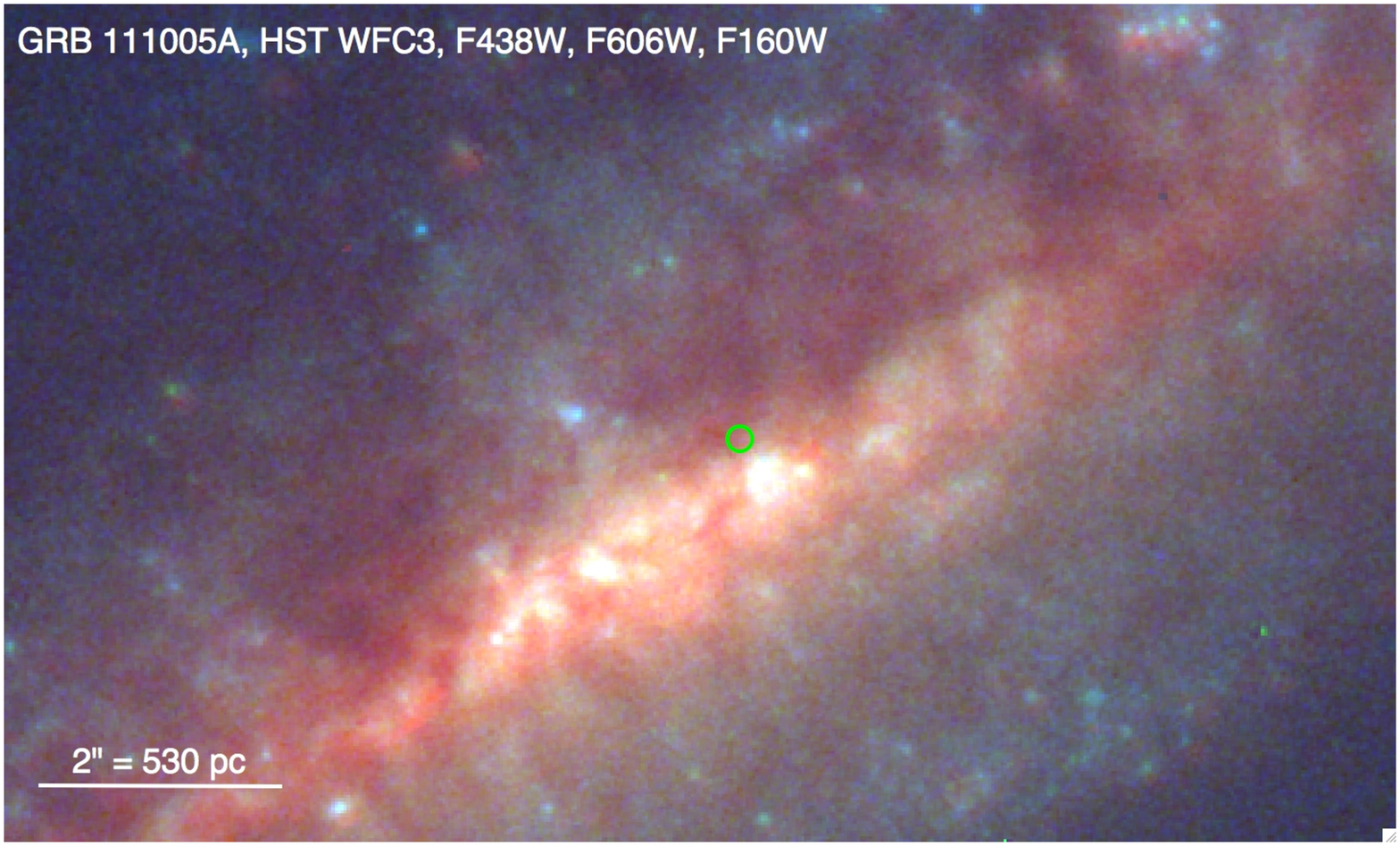}
\end{center}
\caption{HST image of the {\grb} host showing the entire galaxy ({\it left}) and the central region {\it right}. The {\it green circle} shows the {\grb} afterglow VLBA position. It is located behind a dust lane (with suppressed optical emission).}
\label{fig:hst}
\end{figure*}

\begin{figure}
\begin{center}
\includegraphics[width= 0.5\propwidth]{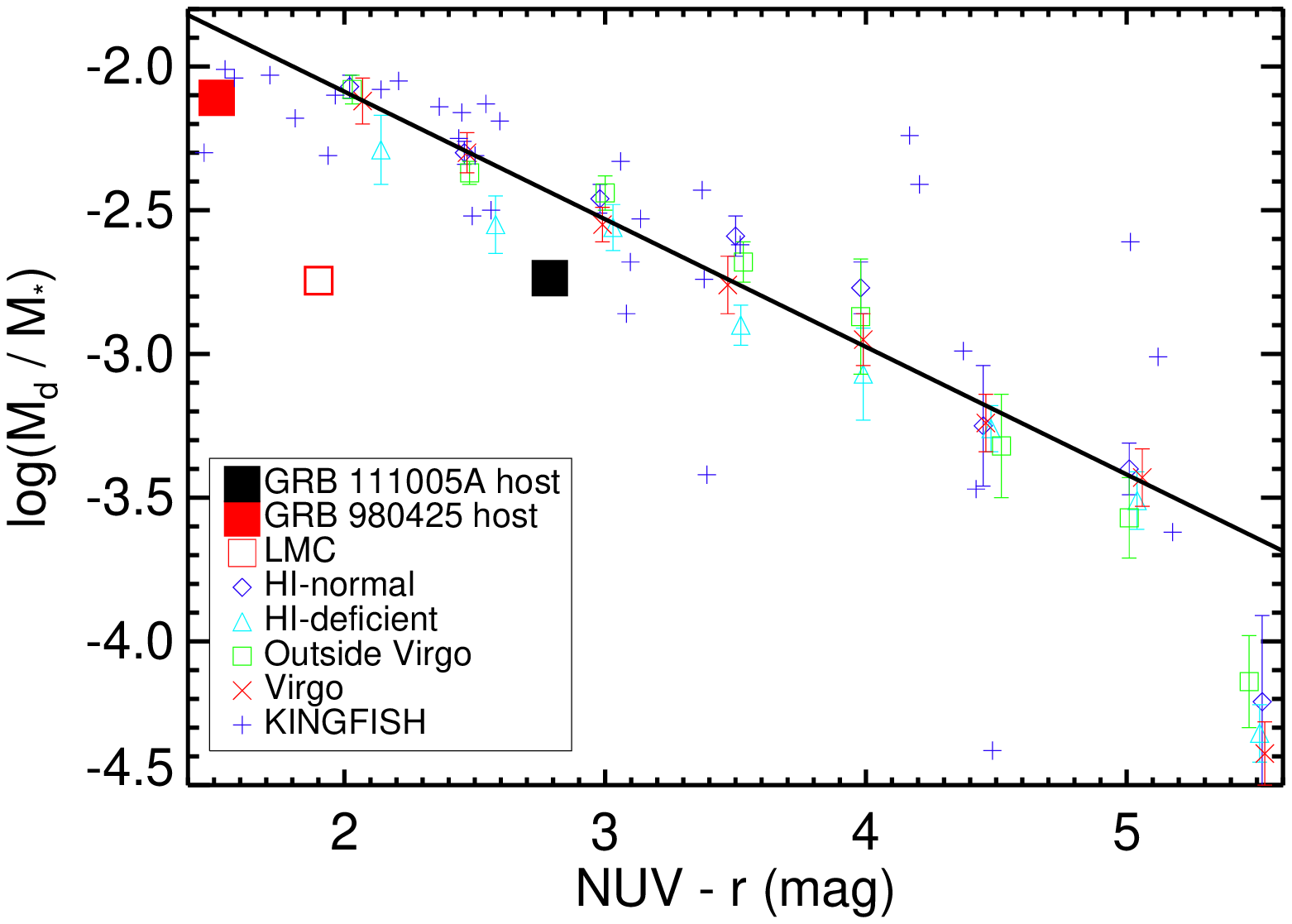}
\end{center}
\caption{Dust-to-stellar mass ratio as a function of UV-to-optical colour of the {\grb} ({\it black square}), GRB 980425 host ({\it red square}; \citealt{michalowski14}), LMC 
({\it open red square}; $M_*$ from \citealt{skibba12} and $M_d$ derived applying the method of \citealt{bianchi13} to the data from \citealt{meixner13}),
KINGFISH galaxies 
\citep[{\it blue plusses};][]{kingfish},
and the averages of other local galaxies in eight colour bins  \citep[table~1 and fig.~4 of][]{cortese12}. The {\it solid line} represents a linear fit to the data. Both GRBs 111005A and 980425 are a factor of $\sim2$ below the trend. This figure is reproduced from \citet{michalowski14}.
}
\label{fig:mdmsnuvr}
\end{figure}

\section{Discussion}
\label{sec:discussion}

\subsection{Association of GRB\,111005A with ESO 580-49 at $z= 0.01326$}
\label{sec:assoc}

The redshift of {\grb} has not been measured from the afterglow emission, so we provide here the evidence that {\grb} exploded in ESO 580-49.  First of all, there is little doubt that the radio object detected by ATCA, WSRT and VLBA before 40 days after the burst is the afterglow of {\grb}. Even though there have been cases of the detection of unrelated variable radio sources near supernova positions \citep[e.g.][]{brunthaler10},  
it is highly unlikely to find a variable decaying radio object visible for only 40 days (with no re-brightening at least till $\sim2\,000$ days) spatially and temporarily coincident with a $\gamma$-ray error circle of a few arcmin. Specifically, the object detected by VLBA must be the afterglow, because otherwise such an $\sim0.7\,$mJy object would be detected by late time ATCA observations at a similar frequency, which have coarser spatial resolution and lower noise (see Table~\ref{tab:after}), so  if the source was not decaying, ATCA would have detected an equal or stronger signal.

Moreover, its radio spectrum rises with frequency (Fig.~\ref{fig:sed}), likely due to self-absorption. This is unheard of for star-forming galaxies or AGN, but is expected for GRB afterglows. 
Hence, we treat the VLBA position as the most precise position of {\grb}.

Moreover, the flux evolution at 18 and 34 \,GHz (Fig.~\ref{fig:lightcurve}) is not due to differences in resolution (and hence different contamination by the host galaxy), because $\sim2\,000$ days after the burst we did not detect any signal using the array configuration similar to the one used for early observations (see Table~\ref{tab:after}), which resulted in detections.
 Indeed, the observations on 2016 Sep 06 and 07 resulted in non-detections even though they have coarser resolution than those on 2011 Nov 02 and 08, during which we obtained strong detections. This would not be the case if the flux drop was the result of the difference in resolution.

At lower frequencies the situation is different. The WSRT detection on 2011 Oct 18 during the EVN run is not consistent with the EVN upper limit. 
This is can be explained if the signal is dominated by the extended emission of the host galaxy, not the point-like emission of the GRB. However, this may also be due to correlation losses during the EVN observations. Even more revealing is the inconsistency between the VLA and ATCA observations at 5.5--5.8\,GHz (Fig.~\ref{fig:lightcurve} and Table~\ref{tab:after}). The VLA observations with poorer spatial resolution resulted in a detection of a non-variable source, whereas the ATCA observations with better resolution resulted in an upper limit approximately three times lower than the VLA detection. This strongly suggests that the emission detected by VLA is related to the host galaxy, not the GRB. We therefore conclude that the emission of the afterglow at the frequencies lower than 10\,GHz was not detected. This may be due to strong self-absorption.

The question of whether  {\grb} is hosted by ESO 580-49, or is a background object is less definite, and must be addressed on statistical grounds. We estimated the chance coincidence of such a bright galaxy 
using the SDSS $r$-band counts \citep[][their table 2]{yasuda01}, showing $\sim0.42\,\mbox{deg}^{-2}$ galaxies brighter than  ESO 580-49 ($r\sim14\,$mag). It is somewhat arbitrary to estimate the angular distance of {\grb} to  ESO 580-49, as it is located inside the galaxy disk. The sub-arcsec VLBA position is $\sim1\arcsec$ from the galaxy center as seen on the $3.6\,\micron$ image (Fig.~\ref{fig:im}), which  gives a negligibly small probability of chance association of $\sim10^{-7}$, so that such a bright galaxy should not be found by chance even in a sample of $\sim1000$ {\it Swift} GRBs. 
This implies that ESO 580-49 is the host galaxy of {\grb}.

Hence, from now on we assume $z= 0.01326$ as a redshift of {\grb}. This is the second closest GRB discovered to date, closely following the GRB 980425 at $z=0.0085$ \citep{tinney98}.

\subsection{Prompt emission}
\label{sec:prompt}

The redshift of $z= 0.01326$ implies a peak luminosity of  $L_\gamma=(6.0\pm1.4)\times10^{46}\,\mbox{erg s}^{-1}$ ($15$--$150$ keV band). 
This is comparable to the peak luminosity of GRB 980425 of $(5.5\pm0.7)\times10^{46}\, \mbox{erg s}^{-1}$ ($24$--$1820$ keV band; \citealt{galamanature}) implying that {\grb} was also a low-luminosity burst. 

 The shape of the lightcurve of the prompt emission of {\grb} \citep[Fig.~\ref{fig:prompt};][]{swiftlc1,swiftlc2} 
was similar in some respects to that of other SN-less long GRBs: 060505 \citep{gehrels06} and 060614 \citep{mcbreen08}. All three GRBs exhibit a prominent peak at the time of the trigger. The difference is that the peak was $\sim5\,$s long for GRBs\,060505 and 060614, whereas it was 26\,s long for {\grb}. Moreover, the peak for GRB\,060614 was followed by a fainter emission lasting $\sim60$\,s, which was not the case for the other two. 

\subsection{Afterglow emission}
\label{sec:afterglow}

{\grb} was also sub-luminous when it comes to its radio emission. Fig.~\ref{fig:lumt} shows that it is an order of magnitude less luminous than the local low-luminosity GRBs. In particular the 18\,GHz emission of {\grb} (green points on this figure) is ten times less luminous than the GRB\,031203 \citep{soderberg04} and 060218 \citep{soderberg06nature} at 22.5\,GHz at a similar epoch. Similarly, 5 and 9\,GHz limits for {\grb} (violet and blue arrows on this figure) are 10--100 times lower than the luminosity of GRB\,980425 \citep{kulkarni98}, 031203, 060218, and GRB\,100316D \citep{margutti13}.

The temporal behaviour of the radio afterglow of {\grb} was also unusual. It shows a relatively constant flux, and then a rapid decay ($\alpha\sim4.5$ with $F\propto t^{-\alpha}$) approximately a month after the GRB explosion. A flat evolution with a subsequent decay is expected from the model of the burst exploding in the uniform medium, but the break should occur after a few days and the decay slope should be shallower  ($\sim2$--$3$; fig.~1 of \citealt{smith05}). Such rapid decay has not been observed for any other GRB \citep[e.g.][]{soderberg04,smith05b, vanderhorst05,vanderhorst14,deugartepostigo12}.

We fitted a line to the afterglow spectral energy distributions in the log-log space (power-law $S_\nu\propto\nu^\alpha$; Fig.~\ref{fig:sed}) and determined the following parameters:
\begin{eqnarray}
\log(S_\nu/\mbox{mJy}) &=& (0.87\pm0.25)\times\log(\nu/\mbox{GHz})  \label{eq:sed1} \\
\nonumber					&& - (0.96\pm 0.36)  \mbox{ at } 4\mbox{--}6 \mbox{ days} \\
\log(S_\nu/\mbox{mJy}) &=& (0.74\pm0.15)\times\log(\nu/\mbox{GHz}) \label{eq:sed2} \\
\nonumber					&& - (0.76\pm 0.19)  \mbox{ at } 12\mbox{--}16.5 \mbox{ days} 
\end{eqnarray}
These lines are consistent with each other within errors, and the slope of $\alpha=0.74$--$0.87$ is within the range of slopes of other GBRs \citep[e.g.][]{smith05b,vanderhorst05,vanderhorst14,deugartepostigo12}. On the other hand, this slope is unlikely for AGN emission, which shows negative slopes (e.g.~fig.~2 of \citealt{prandoni09}), except of blazars (which is not the case for this galaxy, as it is viewed edge-on).

Our VLBA upper limit on the size of the afterglow 16.5 days after the burst of $< 0.38\,$mas corresponds to $<0.1\,$pc at $z= 0.01326$. This is smaller than the size of the radio afterglow of GRB 030329 of $0.19\pm0.06\,$pc measured 24.5 days after the burst \citep{taylor04}. Depending on the model, at $t=16.5$ days these data suggest the size of $0.14$--$0.19\,$pc for GRB 030329, so {\grb} expanded a factor of $>1.5$--$2$ slower. The mean apparent expansion velocity of {\grb} is $<1.1\times10^6\,\kms$ ($<3.7c$; consistent with both mildly relativistic and non-relativistic expansion), indeed lower than $\sim6c$ measured for GRB 030329 at $24.5$ days \citep{pihlstrom07}.

\subsection{The lack of a supernova}

The non-detection of SN emission in the near-IR for {\grb} (Fig.~\ref{snlimits}) cannot be ascribed to dust obscuration. Hence, it effectively rules out the origin of {\grb} as either a classical long-GRB, or any other standard core collapse event down to limits $\sim20$ times fainter than luminosities measured for other SNe associated with GRBs. This is similar to GRB\,060505 and 060614, for which \citet{fynbo06}, \citet{dellavalle06}, and \citet{galyam06} did not detect SN emission despite deep observations (see also  \citealt{gehrels06}). 

The lack of SN emission in long GRBs has been a subject of a long debate \citep{fynbo06,dellavalle06,galyam06,gehrels06,king07,hjorthsn,retter12,yang15,jin15}.
Theoretical stellar explosion models explain such behaviour by  invoking explosions that do not result in the ejection of large amounts of nickel \citep{heger03,fryer06}. This happens either because nickel is not produced, or falls back on the forming black hole. It is likely that {\grb} belongs to this category.

Alternatively, the lack of SN in GRB\,060614, together with a weak near-infrared bump was interpreted as a Li-Paczy\'{n}ski macronova/kilonova resulting from a compact binary merger \citep{yang15,jin15,wang17grb,dado18,yue18}. We do not have necessary near-infrared data to test this hypothesis for {\grb}, but it is also plausible. For future bursts it will be possible to test this with gravitational wave observatories \citep{li16,ligo17,kim17}.

\subsection{Mechanism of explosion}

Based on the result presented above we discuss here all the possible mechanisms for {\grb}: classical core-collapse event, off-axis GRB, AGN activity, tidal disruption event (TDE), and X-ray binaries. We conclude that none of these models explains all the properties of {\grb} fully, so this burst likely represents a new type of explosions not characterised yet before.

\paragraph{Classical core-collapse event.}

The classical long GRB collapsar model is disfavoured mostly by the rapid flux decline ($\alpha\sim4.5$; Section~\ref{sec:prompt}) at $\sim30$ days after the burst.  Moreover, low $\gamma$-ray and radio luminosities, and the lack of a SN suggest that  {\grb} represents a different class of GRBs than typical core-collapse events. 

\paragraph{Off-axis GRB.}

A GRB with the jet axis at an angle to the line-of-sight exhibits a different afterglow evolution \citep{vaneerten10,vaneerten11,kathirgamaraju16}. We explored the off-axis GRB library of \citet{vaneerten10}\footnote{\url{http://cosmo.nyu.edu/afterglowlibrary}}, and found these models inconsistent with our data in three aspects: {\it i}) they do not reproduce sharp flux decline after $\sim30$ days; {\it ii}) they exhibit slightly rising instead of flat evolution before $\sim30$ days; {\it iii}) they exhibit too flat spectral slope compared with our data. 
 Finally, the off-axis GRB model does not explain the lack of a SN for {\grb}.

\paragraph{Active galactic nuclei.}

The VLBA position of {\grb} (with $0.2\,$mas uncertainty) is $\sim1\arcsec$ ($\sim300$\,pc) from the galaxy center where the black hole is likely located, so this scenario is unlikely (Fig.~\ref{fig:hst}). However, it cannot be ruled out that due to the projection effect the supermassive black hole is located behind the dust cloud at the position we detected {\grb}. This would then be the first long GRB associated with an AGN, and the second GRB in general, after the short GRB\,150101B \citep{xie16}.
However, the positive radio spectral slope of {\grb} (Section~\ref{sec:afterglow}) disfavours the AGN scenario.

\paragraph{Tidal disruption event.}

A TDE occurs when a star is torn apart by a supermassive black hole due to tidal forces \citep{hills75,rees88}.  
{\grb} was detected close in the central area of a galaxy which makes this scenario more promising than ever for a GRB event. However, the VLBA position of {\grb} appears to be $\sim1\arcsec$ from the exact galaxy center where the black hole is likely located (Fig.~\ref{fig:hst}).
Moreover, the radio lightcurve of a well-studied TDE was shown to rise over a few hundred days after the onset \citep{bloom11,levan11,zauderer11,zauderer13,berger12}, unlike this GRB. Hence, this scenario is unlikely for {\grb}.

\paragraph{X-ray binary.}

An X-ray binary is a system of a compact object and another star, in which the matter is flowing from the latter to the former. 
The radio behaviour of {\grb} is inconsistent with that of X-ray binaries, because they exhibit nearly flat radio spectra \citep[][compare with Fig.~\ref{fig:sed}]{bogdanov15,tetarenko15}. 
Moreover, the luminosity density of {\grb} of $\sim10^{28}\,\mbox{erg}\,\mbox{s}^{-1}\,\mbox{Hz}^{-1}$ at $18$\,GHz (Fig.~\ref{fig:lumt}) corresponds to the luminosity of $\sim10^{38}\,\mbox{erg}\,\mbox{s}^{-1}$, which is $3$--$4$ orders of magnitudes higher than luminosities of X-ray binaries \citep[e.g.][]{bogdanov15}.

\subsection{Environment of the GRB explosion}

The HST observations are shown in Fig.~\ref{fig:hst}, where wide and narrow fields of view are presented. The observations provide an excellent view of the morphology of the galaxy, which appears to be a slightly disturbed spiral galaxy. Unlike for the vast majority of GRB hosts the high spatial resolution of the HST images resolves the galaxy at the $\sim$ 24 parsec level. This means that numerous individual star forming regions are visible as well as prominent dust lanes. There is little evidence for any prominent bulge. The VLBA position for {\grb} is clearly offset from the nucleus of the galaxy, and lies in a dust lane in a region that does show strong IR emission typical of obscured star forming regions. The position offers direct evidence that GRBs can reside behind significant dust columns, but the absence of any counterpart in the near and mid-infrared  suggests that the dust extinction would have to be extreme to evade detection.

In the VLT/X-shooter spectrum (Fig.~\ref{fig:Xslit}) we, in addition to the continuum from the host, detect strong emission lines. The emission line profile along the slit clearly show two peaks. One peak is aligned with the centre of the host galaxy and the other peak is offset by about $4.5\arcsec$ to the Northwest along the slit. This is illustrated in Fig.~\ref{fig:Xspec} for the [\ion{O}{ii}] and [\ion{O}{iii}] lines. It is clear that the clump offset from the centre of the host has a much harder ionising flux as the [\ion{O}{iii}]/[\ion{O}{ii}] ratio is much stronger here.

\subsection{Global properties of the host galaxy}

ESO 580-49, the host of {\grb}, is clearly a star-forming disk galaxy viewed edge-on with multiple star-forming regions visible in the UV (Fig.~\ref{fig:im}). The galaxy is asymmetric, with the northwest part being more star-forming (or less dust-obscured), as evidenced by more prominent UV emission. On the other hand, near-IR images (tracing the stellar mass distribution) are much more symmetric, and are showing a disk structure of the galaxy.

As shown in Fig.~\ref{fig:sedhost}, at wavelengths longer than $1\,\micron$ the SED of the {\grb} host is very similar to that of the GRB 980425 host (when scaled up by a factor of 4 to match the IR luminosity of the {\grb} host). At shorted wavelengths the scaled GRB 980425 is  brighter, which is partially a consequence of a much lower inclination compared with the nearly edge-on {\grb} host, and hence a much lower dust attenuation. 

Fig.~\ref{fig:mdmsnuvr} shows the dust-to-stellar mass ratio as a function of NUV$-$r colour of the {\grb} and 980425 hosts compared with  other local galaxies observed by {\it Herschel}  (adopted from fig.~5 of \citealt{michalowski14}). It shows that the hosts of both GRBs 111005A and 980425 are close to the lower envelope of the dust-to-stellar mass ratio at their NUV$-$r colours. Recently \citet[][but see \citealt{perley17}]{hatsukade14}, \citet{stanway15}, and \citet{michalowski16,michalowski18} claimed that GRB/SN hosts exhibit low molecular gas masses (see alternative views in \citealt{arabsalmani18} and \citealt{michalowski18co}), and we show here than they may also be dust-deficient. 

As in  \citet{michalowski14} and \citet{kohn15} we integrated the local (mostly $z<0.03$) infrared luminosity function \citep{sanders03} to show that $\sim95$\% of local galaxies are less luminous than {\grb} host  (with $L_{\rm IR}\sim10^{9.6}\,\lsun$), and that $\sim25$\% of the total star formation activity in the local universe happens in these faint galaxies. Having two GRB hosts (980425 and 111005A) below this cut and none above is still consistent with the GRB rate being proportional to the cosmic SFR density (SFRD), but there is a tension with such expectation. 

This can also be demonstrated by calculating a SFR-weighted mean infrared luminosity of local galaxies. If GRBs trace SFRD in an unbiased way their host galaxies should have a mean infrared luminosity similar to this SFR-weighted mean of other galaxies.
 Mean luminosity of galaxies characterised by the luminosity function $\phi$ is $\langle L\rangle=\int_{L_{\rm min}}^{L_{\rm max}} \phi\cdot LdL / \int_{L_{\rm min}}^{L_{\rm max}} \phi dL$. However, in order to compare this to GRB hosts, the mean must be weighted by SFRs, 
 because a galaxy with a higher SFR has a higher probability to host a GRB.
 Then the mean becomes $\langle L\rangle_{\rm SFR}=\int_{L_{\rm min}}^{L_{\rm max}} \phi\cdot\mbox{SFR}\cdot LdL / \int_{L_{\rm min}}^{L_{\rm max}} \phi\cdot \mbox{SFR} dL$. Assuming $\mbox{SFR}\propto L_{\rm IR}$ and using the parameters of the \citet{sanders03} luminosity function this gives $\langle \log(L/L_\odot)\rangle_{\rm SFR}=10.61^{+0.09}_{-0.10}$, or $\langle \mbox{SFR}\rangle_{\rm SFR}=4.1^{+1.0}_{-0.9}\,\msunyr$ using the \citet{kennicutt} conversion and assuming the \citet{chabrier03} IMF (propagating the errors on the luminosity function parameters using the Monte Carlo method). This value is only weakly dependent on the adopted cut-off luminosities, $\log (L_{\rm min}/\lsun)=7$ and $\log(L_{\rm max}/\lsun)=13$, as it is mostly constrained by the shape of the luminosity function close to its knee. This mean SFR is a factor of $\sim10$ higher than the SFRs of the hosts of GRBs 111005A and 980425, so if higher numbers of low-$z$ GRBs are found in such low-luminosity galaxies, then this would mean that the GRB rate is not simply proportional to the cosmic SFRD, and is biased towards low-luminosity galaxies, at least locally. 
 
 This is in line with the result of \citet{perley13,perley15,perley16b}, \citet{vergani15} and \citet{schulze15} that GRB hosts are biased towards less-massive galaxies than what would be expected from the assumption that the GRB rate is proportional to the cosmic SFRD (see also \citealt{boissier13}). \citet{perley16b} explained this effect by the bias of GRBs against galaxies with super-solar metallicities, which corresponds to the aversion to massive galaxies at low-$z$ (as opposed to high-$z$ when this metallicity cutoff is higher than metallicities of most galaxies, even the massive ones).
 On the other hand, \citet{michalowski12grb}, \citet{hunt14}, \citet{schady14}, \citet{kohn15} and \citet{greiner15} argued for GRB host properties to be consistent with those of general population of star-forming galaxies, so this issue requires more investigation.
 
The sample of SN-less GRBs is small, but it is instructive to compare the {\grb} host with the hosts of other two such GRBs.  Both the hosts of {\grb} and 060505 \citep{thone08} have stellar masses of $\sim10^{9.7}\,\msun$, whereas that of GRB\,060614 is much less massive with $10^{8.2}\,\msun$ \citep{castroceron10}. This range is well within that of other long GRBs \citep{savaglio09,castroceron10,perley16b}. Similarly, SFRs of the hosts of {\grb} and 060505 are similar ($\sim0.5\,\msunyr$), whereas that of the GRB\,060614 host is much smaller ($\sim0.02\,\msunyr$). Finally, the hosts of both {\grb} and 060505 are spirals \citep{ofek07,thone08}, whereas the host of GRB\,060614 is a much more compact galaxy \citep{galyam06}. Hence, we conclude that the hosts of {\grb} and 060505 are similar, but that of GRB\,060614 is a much smaller galaxy. However, none of these properties differentiate them from other long GRBs, though the GRB\,060614 is one of the smallest GRB host.
 
 \subsection{Metallicity of the host and the GRB site}

Indeed, this issue is complicated by the discovery of GRB hosts with solar or super-solar metallicities \citep{prochaska09,levesque10b,kruhler12,savaglio12,elliott13,schulze14,hashimoto15,schady15,stanway15b}. This can either be explained if the metallicity cutoff is not strict, but  only decreases the number of high-metallicity GRBs, not ruling them out, or that these particular GRBs represent a different class of events with different physical mechanism, for example a binary system \citep{fryer05,trenti15}.

We measured high metallicity for the {\grb} host, around $1$--$2$ solar (Fig.~\ref{fig:metal}).  A similar conclusion was drawn by \citet{tanga18} using integral field spectroscopy. Hence, despite similar distance and the host luminosity (Fig.~\ref{fig:sedhost}) compared with those of GRB\,980425, the {\grb} host has much higher metallicity. Moreover, the GRB itself did not explode in the most metal-poor region of the galaxy (Fig.~\ref{fig:metal}), which was the case for other low-$z$ GRBs \citep{christensen08,levesque11,thone14,izzo17}.

Super-solar metallicity of {\grb} host and its mass of $\sim5\times10^9\,\msun$ (Tables~\ref{tab:grasilres} and \ref{tab:magphysres}) are inconsistent with the solar metallicity  cutoff and the mass cutoff of $<2\times10^9\,\msun$ proposed by \citet{perley16b} at $z\sim0$. This suggests that {\grb} belongs to a different category than GRBs for which these cutoff values were derived. Those GRBs are all at higher redshifts than {\grb}, and were selected in an optically-unbiased way. Hence, {\grb} belongs likely to a rare class of event, which is not present in an unbiased sample of the order of hundred events.

\section{Conclusions}
\label{sec:conclusion}

Using the 5--345 GHz observations of the afterglow of {\grb} we found that it is located in an edge-on disk galaxy at  $z = 0.01326$, which makes it the second closest GRB known to date. 
 The low $\gamma$-ray and radio luminosities, rapid decay, lack of a SN and super-solar metallicity suggest that  {\grb} represents a different class of GRBs than typical core-collapse events. 
The existence of two local GRBs in low-luminosity galaxies is still consistent with the hypothesis that the GRB rate is proportional to the cosmic SFR density, but suggests that the GRB rate may be biased towards low SFRs. The hosts of both GRBs 111005A and 980425 also exhibit lower dust content that what would be expected from their stellar masses and optical colours.

\begin{acknowledgements}

We thank Joanna Baradziej and our referee for help with improving this paper; the staff at the ATCA, EVN, VLBA, WSRT, PdBI, APEX, VLT, WHT, {\it Spitzer} and HST for scheduling these observations. 
Axel Weiss for help with the APEX observations; 
and Jan Martin Winters for help with the PdBI observations.

M.J.M. and A.-L.T.~acknowledge the support of the National Science Centre, Poland through the POLONEZ grant 2015/19/P/ST9/04010.
M.J.M.~acknowledges the support of the UK Science and Technology Facilities Council, the SUPA Postdoctoral and Early Career Researcher Exchange Program, and the hospitality at the Harvard-Smithsonian Center for Astrophysics and the National Astronomical Observatories, Chinese Academy of Sciences.
This project has received funding from the European Union's Horizon 2020 research and innovation programme under the Marie Sk{\l}odowska-Curie grant agreement No. 665778.
The Dark Cosmology Centre is funded by the Danish National Research Foundation.
H.D.~acknowledges financial support from the Spanish Ministry of Economy and Competitiveness (MINECO) under the 2014 Ramon y Cajal program MINECO RYC-2014-15686.
L.S.~acknowledges the support by National Basic Research Program of China (No. 2014CB845800) and the National Natural Science Foundation of China (No. 11361140349 and No. 11103083).

The Australia Telescope is funded by the Commonwealth of Australia for operation as a National Facility managed by CSIRO. 
The National Radio Astronomy Observatory is a facility of the National Science Foundation operated under cooperative agreement by Associated Universities, Inc. This work made use of the Swinburne University of Technology software correlator, developed as part of the Australian Major National Research Facilities Programme and operated under licence. 
The European VLBI Network is a joint facility of independent European, African, Asian, and North American radio astronomy  institutes. Scientific results from data presented in this publication are derived from the following EVN
project code: RP018. 
e-VLBI research infrastructure in Europe was supported by the European Union's Seventh Framework Programme (FP7/2007-2013) under grant agreement RI-261525 NEXPReS. 
This work has been supported by the European Commission Framework Programme 7, Advanced Radio Astronomy in Europe, grant agreement No. 227290. 
This publication is based on data acquired with the Atacama Pathfinder Experiment (APEX). APEX is a collaboration between the Max-Planck-Institut fur Radioastronomie, the European Southern Observatory, and the Onsala Space Observatory.  
Based on observations carried out with the IRAM Plateau de Bure Interferometer. IRAM is supported by INSU/CNRS (France), MPG (Germany) and IGN (Spain). 
The Westerbork Synthesis Radio Telescope is operated by the ASTRON (Netherlands Institute for Radio Astronomy) with support from the Netherlands Foundation for Scientific Research (NWO). 
Based on observations made with ESO Telescopes at the La Silla Paranal Observatory under programmes 288.D-5004, 088.D-0523, and 090.A-0088. 
The WHT and its override programme are operated on the island of La Palma by the Isaac Newton Group in the Spanish Observatorio del Roque de los Muchachos of the Instituto de Astrofisica de Canarias. 
Based on observations made with the NASA/ESA Hubble Space Telescope, obtained at the Space Telescope Science Institute, which is operated by the Association of Universities for Research in Astronomy, Inc., under NASA contract NAS 5-26555. These observations are associated with program \# 13949. 
This work made use of data supplied by the UK Swift Science Data Centre at the University of Leicester. 
This research has made use of the NASA/ IPAC Infrared Science Archive, which is operated by the Jet Propulsion Laboratory, California Institute of Technology, under contract with the National Aeronautics and Space Administration. 
This research is based on observations with AKARI, a JAXA project with the participation of ESA. 
{\em Galaxy Evolution Explorer} ({\em GALEX}) is a NASA Small Explorer, launched in 2003 April. We gratefully acknowledge NASA's support for construction, operation, and science analysis for the {\em GALEX} mission, developed in cooperation with the Centre National d'Etudes Spatiales of France and the Korean Ministry of Science and Technology. 
This research has made use of data obtained from the High Energy Astrophysics Science Archive Research Center (HEASARC), provided by NASA's Goddard Space Flight Center. 
This publication makes use of data products from the Two Micron All Sky Survey, which is a joint project of the University of Massachusetts and the Infrared Processing and Analysis Center/California Institute of Technology, funded by the National Aeronautics and Space Administration and the National Science Foundation. 
This research was made possible through the use of the AAVSO Photometric All-Sky Survey (APASS), funded by the Robert Martin Ayers Sciences Fund. 
This research has made use of 
the GHostS database (\url{www.grbhosts.org}; \citealt{savaglio09}), which is partly funded by Spitzer/NASA grant RSA Agreement No. 1287913; 
the GRB list maintained by Jochen Greiner (\url{www.mpe.mpg.de/~jcg/grbgen.html});
the NASA/IPAC Extragalactic Database (NED) which is operated by the Jet Propulsion Laboratory, California Institute of Technology, under contract with the National Aeronautics and Space Administration;
SAOImage DS9, developed by Smithsonian Astrophysical Observatory \citep{ds9};
the NASA's Astrophysics Data System Bibliographic Services;
and the  Edward Wright Cosmology Calculator \url{www.astro.ucla.edu/~wright/CosmoCalc.html} \citep{wrightcalc}.

\end{acknowledgements}



\end{document}